\documentclass{aa}

\usepackage{natbib}

\usepackage{graphicx} 
\usepackage{adjustbox}
\usepackage{comment}
\usepackage{amsmath,bm}	
\usepackage{wasysym}    
\usepackage{booktabs}
\usepackage{xspace}
\usepackage[version=4]{mhchem}
\bibpunct{(}{)}{;}{a}{}{,} 
\usepackage{txfonts,textcomp}
\usepackage[colorlinks=true,allcolors=blue]{hyperref}
\usepackage{orcidlink}
\providecommand{\orcid}[1]{\orcidlink{#1}}
\makeatletter 
  \patchcmd{\NAT@citex}
    {\@citea\NAT@hyper@{%
      \NAT@nmfmt{\NAT@nm}%
      \hyper@natlinkbreak{\NAT@aysep\NAT@spacechar}{\@citeb\@extra@b@citeb}%
      \NAT@date}}
    {\@citea\NAT@nmfmt{\NAT@nm}%
    \NAT@aysep\NAT@spacechar\NAT@hyper@{\NAT@date}}{}{}

  \patchcmd{\NAT@citex}
    {\@citea\NAT@hyper@{%
      \NAT@nmfmt{\NAT@nm}%
      \hyper@natlinkbreak{\NAT@spacechar\NAT@@open\if*#1*\else#1\NAT@spacechar\fi}%
        {\@citeb\@extra@b@citeb}%
      \NAT@date}}
    {\@citea\NAT@nmfmt{\NAT@nm}%
    \NAT@spacechar\NAT@@open\if*#1*\else#1\NAT@spacechar\fi\NAT@hyper@{\NAT@date}}
    {}{}
\makeatother

\newcommand\Msun{\text{M}_{\astrosun}} 
\newcommand\Lsun{\text{L}_{\astrosun}} 
\newcommand\Zsun{\text{Z}_{\astrosun}} 

\newcommand\HI{\ion{H}{I}\xspace} 
\newcommand\HII{\ion{H}{II}\xspace} 
\newcommand\HeI{\ion{He}{I}\xspace} 
\newcommand\HeII{\ion{He}{II}\xspace} 
\newcommand\HeIII{\ion{He}{III}\xspace} 
\newcommand\CI{\ion{C}{I}\xspace} 
\newcommand\CII{\ion{C}{II}\xspace} 

\newcommand\OI{\ion{O}{I}\xspace} 
\newcommand\arepo{\textsc{arepo}\xspace}
\newcommand\areport{\mbox{\textsc{arepo-rt}}\xspace}
\newcommand\thesan{\mbox{\textsc{thesan}}\xspace}
\newcommand\thesanone{\mbox{\textsc{thesan-1}}\xspace}
\newcommand\thesandarkone{\mbox{\textsc{thesan-dark-1}}\xspace}
\newcommand\thesanzoom{\mbox{\textsc{thesan-zoom}}\xspace}
\def \kms  {$\rm km\,s^{-1}$\xspace}

\usepackage{xcolor}

\definecolor{m56color}{HTML}{A2D94D}
\definecolor{m55color}{HTML}{009E74}
\definecolor{m47color}{HTML}{9368AB}
\definecolor{m45color}{HTML}{F47F1E}
\definecolor{m44acolor}{HTML}{0073B1}
\definecolor{m44bcolor}{HTML}{859ED7}
\definecolor{m44ccolor}{HTML}{57B4E9}
\definecolor{m36color}{HTML}{C93735}

\newcommand{\colorswatch}[1]{\textcolor{#1}{\rule{1.2em}{0.7em}}}

\begin{document}

   \title{RIGEL: Ultra-faint dwarf galaxy diversity shaped by inhomogeneous cosmic reionization
   }
   \authorrunning{Y. Deng et al.}
   \titlerunning{RIGEL: UFD diversity}


  \author{Yunwei~Deng\inst{1}\orcid{0000-0002-7478-6427}
          \and Hui~Li\inst{1}\orcid{0000-0002-1253-2763}
          \and Ewald~Puchwein\inst{2}\orcid{0000-0001-8778-7587}
          \and Boyuan Liu\inst{3}\orcid{0000-0002-4966-7450}
          \and Aaron~Smith\inst{4}\orcid{0000-0002-2838-9033}
          \and Rahul~Kannan\inst{5}\orcid{0000-0001-6092-2187}
          \and Federico~Marinacci\inst{6,7}\orcid{0000-0003-3816-7028}
          \and Greg~L.~Bryan\inst{8}\orcid{0000-0003-2630-9228}
          \and Kung-Yi~Su\inst{9}\orcid{0000-0003-1598-0083}
          \and Mark~Vogelsberger\inst{10}\orcid{0000-0001-8593-7692}
          }

   \institute{
             1. Department of Astronomy, Tsinghua University, Haidian DS 100084, Beijing, China\\
             \email{\href{mailto:hliastro@tsinghua.edu.cn}{hliastro@tsinghua.edu.cn}}\\
             2. Leibniz-Institut f\"ur Astrophysik Potsdam, An der Sternwarte 16, 14482 Potsdam, Germany\\
             3. Institut f\"ur Theoretische Astrophysik, Zentrum f\"ur Astronomie, Universit\"at Heidelberg, D-69120 Heidelberg, Germany\\
             4. Department of Physics, The University of Texas at Dallas, Richardson, Texas 75080, USA\\
             5. Department of Physics and Astronomy, York University, 4700 Keele Street, Toronto, ON M3J 1P3, Canada\\
             6. Department of Physics \& Astronomy ``Augusto Righi'', University of Bologna, via Gobetti 93/2, I-40129 Bologna, Italy\\
             7. INAF, Astrophysics and Space Science Observatory Bologna, Via P. Gobetti 93/3, I-40129 Bologna, Italy\\
             8. Department of Astronomy, Columbia University, New York, NY 10027, USA\\
             9. CIERA and Department of Physics and Astronomy, Northwestern University, Evanston, IL 60201, USA \\
             10. Department of Physics, Kavli Institute for Astrophysics and Space Research, Massachusetts Institute of Technology, Cambridge, MA 02139, USA
             }

   \date{Draft version \today}

 
  \abstract
   {Ultra-faint dwarf galaxies (UFDs) are among the smallest and oldest galaxies in the Universe and are widely regarded as relics of cosmic reionization. To investigate how reionization quenches star formation and shapes the diversity of UFDs, we present a suite of eight cosmological zoom-in simulations of isolated UFDs with present-day halo masses of $\sim10^9\,\Msun$. The simulations are performed with the radiation-magnetohydrodynamic galaxy formation framework Realistic ISM modeling in Galaxy Evolution and Lifecycles (RIGEL), coupled to realistic large-scale radiation fields extracted from the \thesan reionization simulation. Despite residing in similar $z=0$ halos, the simulated galaxies span nearly two orders of magnitude in stellar mass and broadly reproduce the observed luminosities, sizes, metallicities, and stellar kinematics of Local Group UFDs. We find that reionization quenches star formation through a two-stage process. The arrival of the ionization front rapidly photoionizes the diffuse circumgalactic and intergalactic gas, suppressing further gas accretion onto the galaxy. Star formation nevertheless continues for several hundred Myr using the surviving self-shielded gas reservoir and ceases only after this gas is consumed or dispersed. Within 500 Myr after reionization, less than 40\% of the initial gas mass remains in the halo, with photoevaporation constituting the dominant gas-loss channel. We further show that the halo mass at the time of reionization is a key parameter governing the subsequent evolution of UFDs. Galaxies residing in more massive halos at reionization retain gas for longer periods and undergo more extended chemical enrichment. Consequently, the halo mass at reionization strongly correlates with the final stellar mass, stellar age spread, and chemical evolution of the galaxy. }


   \maketitle
%

\section{Introduction}
Dwarf galaxies are the most numerous and diverse population of galaxies in the Universe. They are considered the fundamental building blocks of massive galaxies in a $\Lambda$CDM framework. At the faintest end, with V-band magnitudes of $M_V\geq-8$, lie the ultra-faint dwarf (UFD) galaxies \citep{2019ARA&A..57..375S:Simon}, which represent the lowest-mass systems still capable of forming stars. Because of their shallow potential wells, UFDs are extremely susceptible to both internal and external feedback processes, making them an ideal setting for investigating how such feedback regulates star formation activity and alters the dark matter (DM) distributions \citep{2013MNRAS.432.1989S:Simpson,2015ApJ...804...18A,2017ApJ...848...85J,2019MNRAS.490.4447W:Wheeler,2022ApJ...941..120G:Gutcke,2023ApJ...959...31K:Kim}. With the advent of next-generation sky surveys, such as the Legacy Survey of Space and Time (LSST) and the primary survey of the China Space Station Telescope (CSST), many UFD galaxies anticipated to be discovered in the local volume and at greater distances will open up new avenues for studying galaxy formation and cosmology \citep[e.g.][]{2023MNRAS.523..876Q:Qu,2025OJAp....8E..89T}.

Observations of UFDs have shown that most of their member stars formed at very early times ($z>6$) \citep{2014ApJ...796...91B:Brown,2014ApJ...789..147W:Weisz,2021ApJ...920L..19S:Sacchi,2025ApJ...992..106D:Durbin}. This implies that the intense background radiation during and after the cosmic reionization era critically shaped the formation and destiny of UFDs. The external ionizing radiation removes the unshielded gas from the galaxy rapidly, and the shallow gravitational potential wells of UFDs suppress the accretion of the intergalactic medium (IGM) after it has been photoionized \citep{2004ApJ...601..666D,2008MNRAS.390..920O,2014MNRAS.444..503N:Noh}. Consequently, the majority of UFDs are quenched during cosmic reionization, with the characteristic halo mass required for reionization-driven quenching typically being below a few times $10^9\,\Msun$ at $z\sim0$ \citep{2008MNRAS.390..920O,2017MNRAS.472.2356M,2022ApJ...941..120G:Gutcke}. The halo growth and star formation histories of UFDs are thus extremely important for setting their present-day observational properties and resulting in the diversity of UFDs located in similar halos \citep{2019ApJ...886L...3R,2026arXiv260222206L:Lin}. 

Cosmic reionization is powered by ionizing photons produced by the earliest galaxies, which carve out growing ionized bubbles within the surrounding IGM. These bubbles overlapped in a spatially inhomogeneous fashion known as ``patchy reionization'' \citep{1986PASP...98.1014S,1996ApJ...461...20H:Haardt,2000ApJ...542..535G:Gnedin}, and by redshift $z \approx 5\text{--}6$ they had come to permeate the entire Universe \citep{2006AJ....132..117F:Fan,2019MNRAS.485L..24K:Kulkarni,2022MNRAS.514...55B:Bosman}. UFDs located in different environments will thus undergo distinct reionization histories and, as a result, produce different observable imprints today. The effects of patchy reionization have been highlighted in observations by the different SFHs of Magellanic and non-Magellanic UFDs \citep{2021ApJ...920L..19S:Sacchi}.

Despite this, most cosmological simulations of dwarf galaxies \citep[e.g.][]{2020MNRAS.491.1656A,2022ApJ...941..120G:Gutcke,2023MNRAS.525.3806M:Martin-Alvarez,2025ApJ...978..129A:Andersson,2025MNRAS.541.1195R:Rey,2025ApJ...986..214G,2026arXiv260222206L:Lin} model the impact of external radiation using a time-dependent but spatially-uniform UV background \citep[UVB; e.g.][]{2009ApJ...703.1416F,2020MNRAS.493.1614F:Faucher-Giguere}. As a result, they do not capture the patchy nature of reionization nor the associated spatial variations in radiation intensity. More importantly, such approximate models typically treat the external radiation merely as an additional ionization and heating source for the exposed gas and correct for the self-shielding of dense regions \citep{2013MNRAS.430.2427R}, whereas in reality, the radiation field and the gas are coupled through intricate radiative transfer and thermochemical processes. Therefore, simulations that seek to capture the detailed formation and evolution of UFDs should aim to incorporate a realistic interaction between galaxies and a spatially inhomogeneous, time-dependent radiation field intensity.

The most accurate approach to model the inhomogeneous, time-dependent radiation field is large-scale galaxy formation simulations with coupled radiation hydrodynamics \citep[RHD;][]{2014ApJ...793...29G:Gnedin,2018MNRAS.479..994R:Rosdahl,2022MNRAS.511.4005K}. Unfortunately, these simulations do not achieve a high enough resolution to examine the UFDs in detail while keeping the computational cost manageable. 
An alternative way is to implement the large-scale radiation field in cosmological zoom-in simulations. \cite{2023ApJ...959...31K:Kim} explored how patchy reionization influences star formation and metallicity in UFDs by employing cosmological zoom-in simulations that incorporate semi-analytically precomputed UV radiation fields. However, they do not explicitly follow the transfer of external radiation, and the semi-analytical model is unable to self-consistently capture the interaction between reionization feedback and galaxy formation. Another approach is adopted by \cite{2024A&A...691A.219B:Baumschlager} and \cite{2025arXiv250819396B:Baumschlager}, which self-consistently models the propagation of UVB generated by discrete sources uniformly distributed in the box with on-the-fly radiative transfer (RT). This method accounts for the spatial and temporal inhomogeneity of the UVB, but its properties depend on the input distribution of sources and therefore lack realistic information about large-scale structures.

The current state-of-the-art zoom-in approach is that of \cite{2025OJAp....8E.153K}, in which the large-scale radiation field is inherited from the \thesan parent box, a large-volume galaxy formation simulation that includes coupled RHD \citep{2022MNRAS.511.4005K}. The radiation flux obtained from \thesan is then projected onto the low-resolution gas cells outside the zoom-in region and serves as the radiation boundary condition for the zoom-in region. These large-scale radiation fields are subsequently followed self-consistently by solving the radiative transfer equations within the zoom-in simulation. This novel method has been successfully implemented in the \thesanzoom simulations to study the high-redshift galaxies in $10^8$ to $10^{13}\,\Msun$ halos with up to $142\,\Msun$ resolution \citep{2025OJAp....8E.153K,2025MNRAS.544..513M:McClymont,2025MNRAS.544..391Z:Zier,2025MNRAS.544..410Z:Zier,2026MNRAS.545f2119S:Shen}.

In this work, we employ this novel radiation boundary technique in the Realistic ISM modeling in the Galaxy Evolution and Lifecycles \citep[RIGEL;][]{2024A&A...691A.231D:Deng} framework and perform a suite of cosmological zoom-in simulations of UFDs. By coupling high-resolution radiation-magnetohydrodynamic galaxy formation simulations to realistic large-scale radiation fields extracted from the \thesan reionization simulation, we self-consistently model the interaction between dwarf galaxies and an inhomogeneous reionization process. Our target is to understand how reionization quenches star formation in UFDs and to investigate whether the diversity observed among Local Group UFDs can arise from variations in their local reionization histories.

The remainder of the paper is organized as follows. In Section~\ref{sec:RIGEL}, we give a brief overview of the RIGEL model. In Section~\ref{sec:sample}, we describe how we select an isolated UFD sample from the \thesan simulation for our zoom-in simulations. In Section~\ref{sec:global_prop}, we report the global properties of the simulated dwarf galaxies. In Section~\ref{sec:reion}, we discuss the impact and imprints of inhomogeneous reionization on the UFDs. Finally, in Section~\ref{sec:discuss}, we discuss and conclude this work.
We adopt the \cite{2016A&A...594A..13P:PlanckCollaboration} cosmology, i.e., $H_0 = 100h\,$km\,s$^{-1}$\,Mpc$^{-1}$ with $h = 0.6774$, $\Omega_\text{m} = 0.3089$, $\Omega_\Lambda = 0.6911$, $\Omega_\text{b} = 0.0486$, $\sigma_8 = 0.8159$, and $n_s = 0.9667$, where all symbols have the usual meaning. All quantities in the paper are reported in physical units (without a scaling factor and the Hubble parameter) unless otherwise specified.

\section{Method}
\label{sec:RIGEL}
\subsection{Gravity, magnetohydrodynamics, and radiative transfer}
\label{sec:solver}
The simulations in this work are performed with the moving-mesh hydrodynamic code \arepo \citep{2010MNRAS.401..791S,2016MNRAS.455.1134P}. Gravity is solved using a hybrid approach that computes the short-range forces with a hierarchical oct-tree method \citep{1986Natur.324..446B,2005MNRAS.364.1105S}, while the long-range forces are calculated using the particle-mesh method \citep{2003gnbs.book.....A}. New features inherited from {\sc gadget-4} \citep{2021MNRAS.506.2871S} are used to improve the performance of gravity calculations, including hierarchical time integration and a correction for correlated force errors at large node boundaries due to nearly static particle distributions at high redshift.

Hydrodynamics is handled using a quasi-Lagrangian finite-volume method, which employs an unstructured moving mesh generated through Voronoi tessellation of discrete mesh-generating points. These mesh-generating points drift as they follow the local gas velocity, ensuring an accurate representation of the fluid dynamics. We employ an explicit (de)refinement scheme for hydrodynamics, in which a cell is refined by splitting it into two halves if its mass exceeds twice the target mass, and derefined by merging it with its neighbors if its mass falls below half the target mass. The ideal MHD equations are solved using an eight-wave Powell cleaning approach in order to enforce the divergence constraint on the magnetic field \citep{2013MNRAS.432..176P:Pakmor}.

The radiation field is explicitly modeled and coupled to gas hydrodynamics by the moment-based RHD solver \areport \citep{2019MNRAS.485..117K}. The radiation from stars is modeled using seven spectral bands, including
infrared (IR, $0.1-1$\,eV), optical (Opt., $1-5.8$\,eV), far-ultraviolet (FUV, $5.8-11.2$\,eV), Lyman--Werner (LW, $11.2-13.6$\,eV), hydrogen ionizing (EUV1, $13.6-24.6$\,eV), \HeI ionizing (EUV2, $24.6-54.4$\,eV), and \HeII ionizing (EUV3, $54.4-\infty$\,eV) bands. The radiative transfer equations are solved by combining its zeroth and first moments with the M1 closure relation \citep{1984JQSRT..31..149L}. To avoid extremely small time steps, we reduce the speed of light to an effective value of $1000$\,\kms \citep{2001NewA....6..437G} inside the high-resolution zoom-in regions. Ionization feedback from \HII regions is also corrected in low-resolution cases using the method outlined by \cite{2024MNRAS.527..478D}.

\subsection{Cooling, heating, and chemistry}
The gas cooling and heating are coupled with the radiation fields via a non-equilibrium thermochemical network, comprehensively outlined in Section~2.2 of \cite{2024A&A...691A.231D:Deng}. Briefly, it tracks the non-equilibrium abundance of \text{\ce{H2}, \HI, \HII, \HeI, \HeII, \HeIII} and models the primordial cooling from hydrogen and helium species ($\Lambda_\text{pri}$), tabulated cooling
rates for high-temperature ($\gtrsim10^5$\,K) metals ($\Lambda_\text{Z,CIE}$), nebular lines cooling in photoionized gas ($\Lambda_\text{Z,neb}$), equilibrium metal cooling in warm and cold gas ($\Lambda_\text{Z,C/O}$, including the cooling by \CI, \CII, \OI, and \ce{CO} lines), cooling due to dust--gas--radiation interaction ($\Lambda_\text{dust}$), and Compton cooling of the cosmic microwave background ($\Lambda_\text{CMB}$). The net cooling rate ($\Lambda_\text{net}$) is then given by
\begin{align}
\Lambda_\text{net}&=\Lambda_\text{pri}(n_j, N_\gamma^i,T)+\frac{Z}{\Zsun}\Lambda_\text{Z,CIE}(T,\rho,z)\,\notag\\
&+\Lambda_\text{Z,C/O}(n_j,n_\text{C/O},N_\gamma^{\text{LW}},T)\,\notag\\
&+\Lambda_\text{Z,neb}(n_\text{\HII},n_e,T, Z_\text{g})+\Lambda_\text{dust}(\rho, T, N_\gamma^{\text{IR}},Z_\text{d}) \notag
\\&+\Lambda_\text{CMB}(\rho,T,z)\,,
\end{align}
where $n_j$ is the number density of all the
primordial species $j\in\{\text{\ce{H2}, \HI, \HII, \HeI, \HeII, \HeIII}\}$, $N_\gamma^i$ is the photon number density of all the photon bins from IR to \HeII ionizing bands, $T$ is the gas temperature, $\rho$ is the density of the gas cell, $z$ is the redshift, and $Z$, $Z_\text{g}$, and $Z_\text{d}$ are the total, gas-phase, and dust-phase metallicity, respectively.

\subsection{Star formation and feedback}
\label{sec:SFmodel}
The star formation and feedback are tracked by the RIGEL model \citep{2024A&A...691A.231D:Deng}. Stars are formed in cold ($T <
T_\text{th}$), dense ($n_\text{H} > n_\text{th}$), contracting ($\nabla\cdot {\bm v} < 0$), self-gravitating ($[||\nabla {\bm v}_i||^2+(c_{s,i}/\Delta x)^2]/8\pi G\rho_i<0.5$, where $||\nabla {\bm v}_i||$ is the Frobenius norm of velocity), and marginally Jeans-resolved gas, where the thermal Jeans length is smaller than 4 times the cell size ($L_\text{J}=\sqrt{\pi c_s^2 / G \rho}<4\Delta x$). In this work, given the mass resolution of $17.8\,\Msun$, the density and temperature thresholds for star formation were set as $T_\text{th}=100$\,K and $n_\text{th}=3000$\,cm$^{-3}$, respectively. A gas cell eligible for star formation is converted into a star particle with a probability of ${\cal P}_\text{SF}=\epsilon_\text{ff}\Delta t/\tau_\text{ff}$, where the free-fall time of the cell is given by $\tau_\text{ff}=(3\pi/32G\rho)^{1/2}$, $\Delta t$ denotes the simulation time-step, and a constant star formation efficiency per free-fall time $\epsilon_\text{ff}=1$ is adopted as the resolution is high enough to resolve the fragmentation of star-forming clouds.

Individual massive stars are drawn from the \cite{2003PASP..115..763C} initial mass function (IMF) in the star-forming ISM, and we do not adopt a separate Pop~III IMF (see discussions in Section~\ref{sec:cav}). They provide feedback through radiation, stellar winds, and core-collapse supernovae (SNe). Lifetimes, photon production rates, mass-loss rates, and wind velocities of massive stars (stellar mass $M_\star > 8\,\Msun$) are determined by their initial masses and metallicities based on a library that incorporates a variety of stellar models. This stellar feedback model combines the stellar models covering the metallicity from $10^{-8}\,\Zsun$ to solar from \cite{Schaerer02}, \cite{2003ApJS..146..417L}, \cite{2019MNRAS.482.1304E}, \cite{2020MNRAS.495.4170T}, and \cite{Gessey-Jones2022}, and the details were described in Section~2.4 of \cite{2024A&A...691A.231D:Deng}. We assume that all stars with initial masses larger than $8\,\Msun$ explode as core-collapse SNe and release $10^{51}$\,erg of thermal energy into their neighboring gas cells at the end of their main-sequence lives.

Dying stars release mass and metals through three channels: asymptotic giant branch (AGB) winds, core-collapse SNe, and Type Ia SNe (SN Ia). We use the AGB yield tables from \cite{2010MNRAS.403.1413K}, complemented by the results of \cite{2014MNRAS.441..582D} and \cite{2014ApJ...797...44F}; see \citealt{2018MNRAS.473.4077P} for a detailed description. The yield table for massive stars is adopted from the {\tt set M} of \citet[LC18]{2018ApJS..237...13L}, which covers the mass range of $13-120\,\Msun$. These tables include the final integrated yields and the yields present in the stellar winds. The SNe Ia are modeled using a delay-time distribution (DTD) formalism to compute the explosion probability of each star particle as a function of time. This phenomenological model describes the probability of SN Ia by three parameters: the number of SNe Ia per unit stellar mass $N_\text{Ia}$
formed, the lower bound $\tau_\text{Ia}$, and the slope $s$ of the power-law DTD $\Psi \propto t^{-s}$. The integration of $\Psi(t)$ over a Hubble time $t_\text{H}\approx13.7$\,Gyr equals 1. Here we adopt $s=1.12$ and $N_\text{Ia}=1.3\times10^{-3}$\,SN\,$\Msun^{-1}$ \citep{2012MNRAS.426.3282M:Maoz,2013MNRAS.436.3031V}. $\tau_\text{Ia}\approx40$\,Myr is the lifetime of a $8\,\Msun$ massive star.

As the gas resolution is lower than that of previous non-cosmological simulations using RIGEL, we perform detailed experiments to check whether the Sedov--Taylor phase of SNe is properly resolved. We found that only $1\%$ of SNe do not satisfy the ST-resolving criterion of $m_\text{gas,SN}\,/\,M_\text{shell}<1/27$ by \protect\cite{2015ApJ...802...99K}, where $m_\text{gas,SN}$ is the mass of the gas cell that received the SN energy and $M_\text{shell}$ is the total swept-up mass at the end of the energy-conserving ST phase.

Additionally, we found that the sound speed of the gas cell that received the SN energy ($c_\text{s,SN}$) can exceed the reduced speed of light when the SNe are highly clustered. In this regime, the hydrodynamic timestep required after the SN explosion can become substantially shorter than the timestep assigned to the cell before the energy injection and cannot immediately respond to the sudden increase in signal speed. We thus employed a timestep limiter for the nearest neighboring cells of upcoming SNe following the method outlined by \cite{2021MNRAS.501.5597G}. The timestep size must be smaller than $\Delta t = \min(\Delta x/c_\text{s,SN}, t_\text{SN}-t_\text{now})$ to ensure that the SN is temporally resolved (see also \citet{2025ApJ...987...86H:Hirashima} for the problems of temporal resolution).

\subsection{Large-scale external radiation fields}
In this work, we perform cosmological zoom-in simulations to resolve the smallest dwarf galaxies with very high resolution. However, zoom-in simulations lack the capability to capture the astrophysical processes associated with large-scale structure formation and, as a result, they provide no information about the external large-scale radiation field beyond the zoom-in region.  To model this important ingredient for dwarf galaxy formation, we take advantage of selecting zoom-in galaxies from the parent \thesan simulation. 

\thesan \citep{2022MNRAS.511.4005K,Garaldi2022,2024MNRAS.530.3765G:Garaldi,Smith2022} is a large-volume galaxy formation simulation with fully-coupled RHD, leveraging the IllustrisTNG framework \citep{Weinberger2017,Pillepich2018}. \thesan records the radiation field with a high time cadence (every $\approx 2\text{--}3$\,Myr) by depositing the gas quantities onto a $1024^3$ Cartesian grid, enabling convenient and faithful lookup at any given position and time before redshift 5.5 in the $95.5$\,Mpc box. We model the patchy, time-varying large-scale radiation field based on the \thesan simulation, following the method outlined by \cite{2025OJAp....8E.153K}. We map the radiation flux from \thesan onto the low-resolution gas cells outside the zoom-in region, and we track how this radiation propagates across the boundary of the high-resolution domain to guarantee that the incoming radiation present in the parent \thesan simulation is correctly represented. Meanwhile, the radiative transfer of outgoing radiation produced by local sources within the high-resolution region is followed directly and modeled self-consistently. The \thesan simulation is terminated at $z=5.5$. Between $z=5.5$ and $z=5$, we gradually transition the UVB from that of the final \thesan snapshot to a spatially-uniform UVB by \cite{2009ApJ...703.1416F}, and for redshifts below this range, we adopt this uniform UVB outside the high-resolution zoom-in region.

\subsection{Initial conditions}
We select zoom-in regions from the $z=0$ snapshot of the \thesandarkone simulation in the \thesan project \citep{2022MNRAS.511.4005K,Garaldi2022,2024MNRAS.530.3765G:Garaldi,Smith2022}. This parent simulation is a DM-only simulation of the same initial condition as the \thesan flagship simulation (\thesanone), but it was run to $z=0$, which has a box size of $95.5$\,cMpc and a dark matter resolution of $3.7\times10^6\,\Msun$. The zoom-in isolated dwarf galaxies are found based on the isolation parameter $I$ outlined by \cite{2020MNRAS.491.1656A}. To calculate $I_n$ for a halo $n$, we first calculate the relative distance $D_{n,m}$ between halo $n$ and all more massive halos $m$, where $D_{n,m}=|{\bm r}_n-{\bm r}_m|-R_{200,n}-R_{200,m}$. We then normalize $D_{n,m}$ by $I_{m,n}=D_{n,m}/R_{200,n}$, and define $I_n=\min(I_{m,n})$. We identify halos as isolated when their isolation parameter satisfies $I > 10$.

The initial conditions of the selected halos are constructed using a new zoomed-initial conditions code, \textsc{CosmoZoomIC} (Puchwein et al. in prep), which makes our runs numerically more efficient by optimizing the high-resolution region and the distribution of boundary particles. The resolution levels of the latter can have arbitrary shapes and are chosen based on the distance to the nearest edge of the high-resolution region. This allows us to concentrate the numerical effort more efficiently on the region that is relevant for the scientific analysis. We have also increased the accuracy of the displacement calculation by making the matching in $k$-space between low- and high-resolution power more accurate and using the appropriate Nyquist frequency as $k$-cutoff separately for each resolution level of the boundary particles (Puchwein et al. in prep). We zoom in by six levels from the parent \thesan simulation, leading to a baryon resolution of $17.8\,\Msun$ and DM resolution of $95.2\,\Msun$. This zoom-in level corresponds to the 32x level in the notation of \thesanzoom project \citep{2025OJAp....8E.153K} and an effective particle number of $2\times64200^3$.

To avoid contamination by low-resolution particles, the selected high-resolution region at $z=0$ extends out to 5.5$R_{200}$ following the empirical criterion given by \cite{2014MNRAS.437.1894O:Onorbe}, where $R_{200}$ is the virial radius of the halo. All the simulations presented in this work have zero contamination within the virial radius of the target halo.

In contrast to several recent cosmological zoom-in simulations \citep{2019MNRAS.490.4447W:Wheeler,2020MNRAS.491.1656A,2021ApJ...906...96A:Applebaum}, we do not impose an initial metallicity floor (e.g.\ $10^{-4}\,\Zsun$ or $10^{-3}\,\Zsun$) to mimic pre-enrichment by Pop~III stars. Instead, we assume an initial metallicity of $10^{-8}\,\Zsun$ solely to ensure numerical stability when solving the thermochemistry equations.

\section{Isolated ultra-faint dwarf galaxy selection}
\label{sec:sample}
To study the diversity of isolated UFDs, we consider only the halos with similar $z=0$ mass ($M_{\text{halo}}^{z=0}$) in the range $1.0<M_{\text{halo}}^{z=0}/10^9\,\Msun<1.1$. Following \cite{2020MNRAS.491.1656A}, we define the halos with isolation parameter $I_n>10$ at $z=0$ as isolated halos. In \thesandarkone, we found 29,595 isolated halos in our target mass range, which amounts to $39\%$ of all 75,059 halos in the same mass range. 

After selecting the isolated halos at $z=0$ in the DM-only simulation, we match their bijective cross-linked subhalos in the \thesanone flagship simulation at $z=5.5$ using the catalog compiled by \cite{Zhao2026}. By doing this, we can extract the redshift of reionization ($z_\text{reion}$) and halo mass at $z_\text{reion}$ ($M_\text{halo}^\text{reion}$). Here, \cite{Zhao2026} define the reionization redshift of a halo as the moment when the volume-weighted ionization fraction within a $(125\,\text{kpc})^3$ region enclosing the target halo first reaches 50\%. As $z_\text{reion}$ is evaluated at the halo locations tracked by the merger tree, it is more realistic than most IGM studies that adopt a co-spatial $z_\text{reion}$. These two parameters provide crucial information about the history of the halos that affect the formation and evolution of the diverse UFDs in $10^9\,\Msun$ halos: $z_\text{reion}$ for environmental reionization history, $M_\text{halo}^\text{reion}$ for local halo growth history. We note that at least 20 resolution elements with mass $3.12\times10^6\,\Msun$ are required to identify a halo in the SUBFIND algorithm. Thus, the halos with $M_\text{halo}^\text{reion}<6.24\times10^7\,\Msun$ will be filtered out from our sample. As a consequence, our sample is biased toward the relatively fast-growing halos. In both the all and the isolated halo samples, halos with well-defined $z_\text{reion}$ and $M_\text{halo}^\text{reion}$ account for $\sim60\%$ of the total population.

\begin{figure}
	\includegraphics[width=\columnwidth]{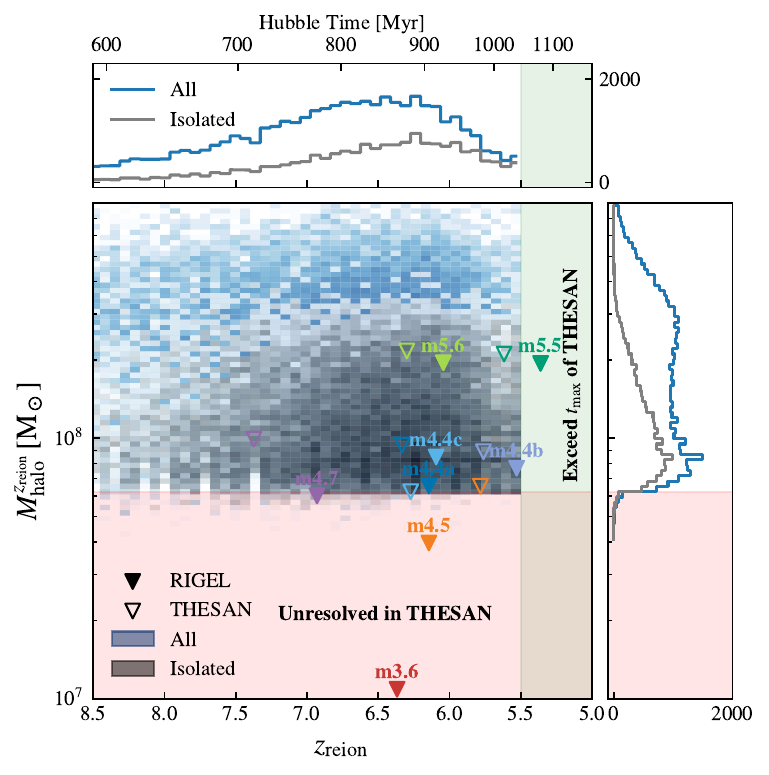}
    \caption{Joint probability distribution of the redshift of reionization ($z_\text{reion}$) and halo mass at $z_\text{reion}$ ($M_\text{halo}^\text{reion}$) of $1.0<M_{\text{halo}}^{z=0}/10^9\,\Msun<1.1$ halos in \thesandarkone, and the corresponding marginalized distributions. The blue histograms represent the distribution of all halos for which both $z_\text{reion}$ and $M_\text{halo}^\text{reion}$ are well defined, whereas the grey histograms are isolated halos with an isolation parameter of $I>10$. Halos with $M_\text{halo}^\text{reion}\lesssim 6.24\times10^7\,\Msun$ are not presented here, because the SUBFIND algorithm requires at least 20 resolution elements of mass $3.12\times10^6\,\Msun$ to identify a halo and to obtain their $z_\text{reion}$ and $M_\text{halo}^\text{reion}$. The hollow triangles mark the $z_{\rm reion}$ and $M_\text{halo}^\text{reion}$ of seven halos in our sample with $M_\text{halo}^\text{reion}> 6.24\times10^7\,\Msun$, extracted from the \thesanone flagship simulation. The corresponding filled triangles show the $z_{\rm reion}$ and $M_\text{halo}^\text{reion}$ obtained in our RIGEL simulations, including the randomly selected halo (m3.6) with $M_\text{halo}^\text{reion}< 6.24\times10^7\,\Msun$.
    }
    \label{fig:thesan}
\end{figure}

In constructing the simulation suite, our goal is to choose representative isolated halos distributed across the $M_\text{halo}^\text{reion}$--$z_\text{reion}$ parameter space. Figure~\ref{fig:thesan} indicates that halos with $M_{\text{halo}}^{z=0}=10^9\,\Msun$ begin to be influenced by reionization when the masses of their progenitors reach $10^8\,\Msun$. Thus, we select three halos that have a similar $M_\text{halo}^\text{reion}$ of $\sim10^8\,\Msun$ but are affected by reionization at redshifts 7.4, 6.3, and 5.8, respectively. To further sample the range in $M_\text{halo}^\text{reion}$, we choose two halos with larger $M_\text{halo}^\text{reion}$ of $\sim2\times10^8\,\Msun$ and another two halos with masses near the $M_\text{halo}^\text{reion}$ halo mass resolution limit of THESAN, $M_\text{halo}^\text{reion}\approx6\times10^7\,\Msun$. Additionally, we randomly selected an $M_{\text{halo}}^{z=0}\approx10^9\,\Msun$ halo below the $M_\text{halo}^\text{reion}$ resolution limit of THESAN. We name the selected galaxies by their present-day stellar mass (details on the naming convention are given in Section~\ref{sec:global_prop}).

In Fig.~\ref{fig:thesan}, we plot the distribution of our eight halos on the $M_\text{halo}^\text{reion}$--$z_\text{reion}$ plane, except for the randomly selected one, as hollow triangles. However, it should be noted that the reionization redshift $z_\text{reion}$ in \cite{Zhao2026} is calculated as an average of a box much larger than the typical virial radius of our selected dwarf halos. Thus, the actual reionization redshift of each re-simulated zoom-in halo can deviate from the value in the catalog. We show the $z_\text{reion}$ and $M_\text{halo}^\text{reion}$ obtained in our RIGEL simulations in Fig.~\ref{fig:thesan} by filled triangles. Overall, the halos in RIGEL tend to be reionized at later times and have lower $M_\text{halo}^\text{reion}$ values, with the exception of the m4.4c galaxy, which is the only case that exhibits a higher $M_\text{halo}^\text{reion}$. The $M_\text{halo}^\text{reion}$ of the randomly selected galaxy can also be obtained in the RIGEL simulation, and it is $\sim10^7\,\Msun$. 

The deviation can arise both physically and numerically: first, the ionization front (I-front) takes different amounts of time to propagate to different positions within the box; second, small-scale turbulent structure changes the speed of I-front propagation; third, we use a reduced speed of light of $1000$\,km\,s$^{-1}$, which is much smaller than the value used in the \thesan simulation. To quantify the timing of reionization consistently, we define the actual reionization redshift $z_\text{reion}$ of an individual halo as the time when half of the gas mass within the $2R_{200}<r<3R_{200}$ shell is ionized. In Appendix~\ref{sec:def_zion} and \ref{sec:rsla}, we discuss the effect of the shell radius and reduced speed of light.

In summary, our sample covers a 300\,Myr range in the halo reionization time ($z_\text{reion}$) and a one order of magnitude range in $M_\text{halo}^\text{reion}$. We introduce the global properties of the simulated UFDs in the following section.

\section{Global properties of the simulated galaxies}
\label{sec:global_prop}
\begin{figure*}
	\includegraphics[width=2\columnwidth]{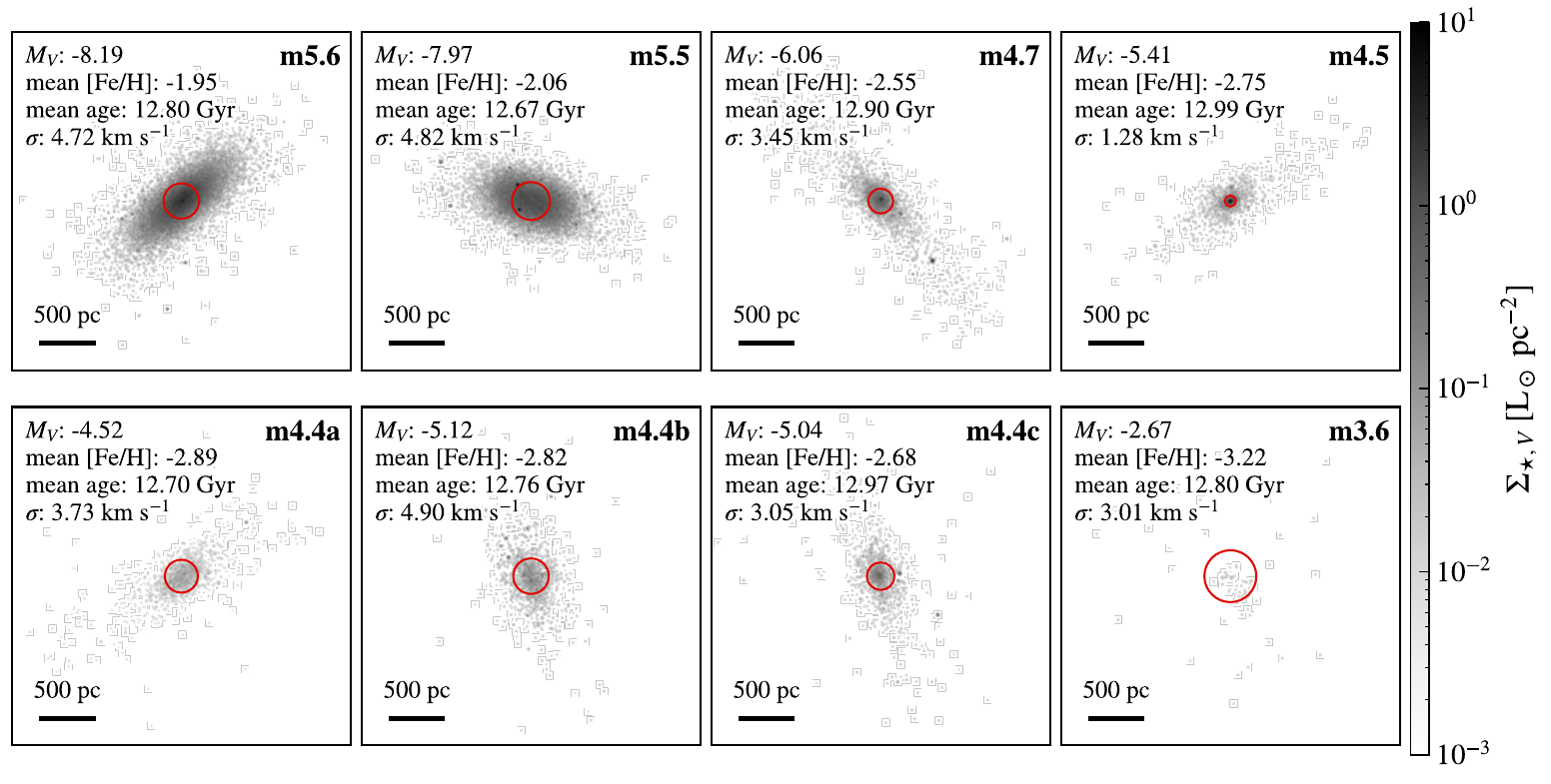}
    \caption{Projected $V$-band surface-brightness maps of the eight RIGEL galaxies at $z=0$.
    Star particles are assigned their $V$-band luminosities and smoothed with a Gaussian PSF of width $\sigma=0.1$~pc (unresolved at the plotted pixel scale of $\simeq 3$~pc).
    Red circles mark the projected $V$-band half-light radius $R_{1/2}$.
    Each panel lists the absolute magnitude $M_V$, the luminosity-weighted mean stellar metallicity and age, and the line-of-sight stellar velocity dispersion $\sigma$. Except for the smallest one, m3.6, the other seven galaxies behave like classic elliptical UFDs. The faintest system (m3.6) is a diffuse ultra-faint dwarf with only $\sim 70$ star particles within $R_{1/2}$.
    }
    \label{fig:Vband}
\end{figure*}
Figure~\ref{fig:Vband} shows the stellar surface brightness distributions of our simulated UFDs at $z=0$. As expected, all eight simulated galaxies present as relics of reionization, with old stellar populations and no gas content. Except for the smallest one, the other seven galaxies behave like classic elliptical UFDs. As these galaxies have similar $z=0$ halo masses, we name them by their present-day stellar mass. For instance, m5.6 indicates a galaxy with a stellar mass of $\log{(M_\star/[\Msun])} = 5.6$. Since there are three galaxies with comparable stellar masses of $\log{(M_\star/[\Msun])} \approx 4.4$, we label them m4.4a, m4.4b, and m4.4c, ordered such that $M_\star(\text{m4.4a}) > M_\star(\text{m4.4b}) > M_\star(\text{m4.4c})$. These three galaxies constitute an interesting subset of galaxies with both similar halo masses and stellar masses. In Table~\ref{tab:properties}, we summarize the properties of all dwarfs.

The lowest-mass galaxy (m3.6) ends up with a final stellar mass of only $3800\,\Msun$, while the two most massive systems each exceed $3\times10^5\,\Msun$ in stellar mass. These results highlight the substantial diversity of galaxies that reside in similar halos but have experienced different evolutionary histories. We note that the most massive galaxy, with $M_V=-8.19$, is not a UFD under the standard criterion \citep[$M_V>-8$,][]{2019ARA&A..57..375S:Simon}, whereas the second most massive one, with $M_V=-7.97$, represents the most massive system that still meets the UFD definition. In contrast, the smallest galaxy, m3.6, is made up of only a few widely scattered star particles, which are barely discernible in Fig.~\ref{fig:Vband}. Such diffuse systems are difficult to identify in observations because they are significantly contaminated by foreground stars. Nonetheless, it still qualifies as a galaxy: a stellar system confined within a dark matter gravitational potential. For simplicity, we still refer to all of these systems as UFDs. 
\begin{table*}
\caption{Properties of the simulated galaxies in the suite. The first two
columns give (1) the names of the galaxies and (2) the colors used to
represent them throughout the paper. The remaining columns give:
(3) $M_{200}$: mass within the radius enclosing a density 200 times the
critical density; (4) $M_{\star}$: stellar mass; (5) $R_{1/2}$: projected
V-band half-light radius; (6) $\sigma$: velocity dispersion of star particles
inside the radius of peak stellar circular velocity; (7) $e_\text{3D}$:
luminosity-weighted intrinsic ellipticity of star particles; (8) $e_\text{2D}$:
projected ellipticity with the 16th--84th percentile range of 30 viewing directions; (9) [Fe/H]:
average stellar iron abundance; (10) $\tau$: average stellar age;
(11) $z_\text{reion}$: reionization redshift; (12)
$M_{200}^\text{reion}$: $M_{200}$ at $z_\text{reion}$; and
(13) $M_{\star}^\text{reion}$: stellar mass at $z_\text{reion}$.}
\centering
\begin{tabular}{ccccccccccccc}
\toprule Name & Color & $M_{200}$ & $M_{\star}$ & $R_{1/2}$ & $\sigma$ & $e_\text{3D}$ & $e_\text{2D}$ & [Fe/H] & $\tau$ & $z_\text{reion}$ & $M_{200}^\text{reion}$ & $M_{\star}^\text{reion}$\\ Unit & -- & [$10^8\Msun$] & [$10^3\Msun$] & [pc] & [km s$^{-1}$] & -- & -- & -- & [Gyr] & -- & [$10^8\Msun$] & [$10^3\Msun$]\\ \midrule m5.6 & \colorswatch{m56color} & 9.02 & 382.8 & 156.5 & 4.72 & 0.57 & $0.50_{-0.18}^{+0.04}$ & $-1.95$ & 12.80 & 6.04 & 1.32 & 159.9\\ m5.5 & \colorswatch{m55color} & 8.60 & 311.6 & 168.4 & 4.82 & 0.51 & $0.38_{-0.15}^{+0.06}$ & $-2.06$ & 12.67 & 5.36 & 1.93 & 169.0\\ m4.7 & \colorswatch{m47color} & 8.17 & 55.6 & 110.0 & 3.45 & 0.76 & $0.57_{-0.17}^{+0.12}$ & $-2.55$ & 12.90 & 6.93 & 0.50 & 14.9\\ m4.5 & \colorswatch{m45color} & 6.34 & 31.3 & 49.5 & 1.28 & 0.70 & $0.48_{-0.09}^{+0.10}$ & $-2.75$ & 12.99 & 6.14 & 0.39 & 4.95\\ m4.4a & \colorswatch{m44acolor} & 8.25 & 23.9 & 145.8 & 3.73 & 0.70 & $0.58_{-0.14}^{+0.05}$ & $-2.89$ & 12.70 & 6.15 & 0.42 & 0.14\\ m4.4b & \colorswatch{m44bcolor} & 7.90 & 23.6 & 157.5 & 4.90 & 0.52 & $0.32_{-0.10}^{+0.08}$ & $-2.82$ & 12.76 & 5.53 & 0.76 & 7.0\\ m4.4c & \colorswatch{m44ccolor} & 7.50 & 22.5 & 122.2 & 3.05 & 0.44 & $0.34_{-0.10}^{+0.07}$ & $-2.68$ & 12.97 & 6.09 & 0.53 & 14.7\\ m3.6 & \colorswatch{m36color} & 7.98 & 3.8 & 230.0 & 3.01 & 0.69 & $0.39_{-0.05}^{+0.18}$ & $-3.22$ & 12.80 & 6.36 & 0.05 & 0.0\\ \bottomrule
\end{tabular}
\label{tab:properties}
\end{table*}

\subsection{Star formation histories}
\label{sec:SFH}
\begin{figure}
	\includegraphics[width=\columnwidth]{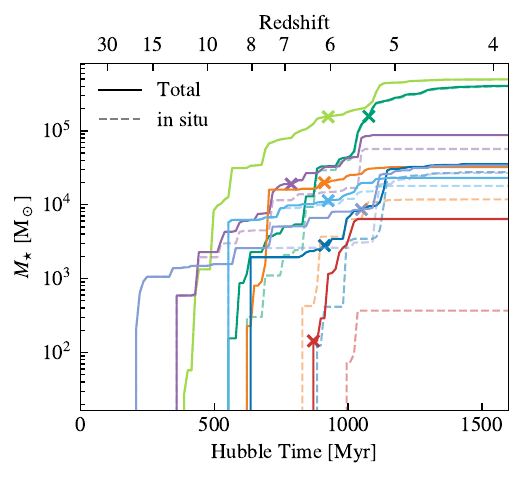}
    \caption{Cumulative star formation history of the simulated dwarf galaxies, shown with the same color scheme as Fig.~\ref{fig:thesan}. The solid curves show the formation history of all stars in the present-day galaxy, while the dashed curves show the histories of the in-situ star formation of the most massive progenitor halos of merger trees. The crosses mark the time when the halo is reionized.
    }
    \label{fig:SFH}
\end{figure}
In Fig.~\ref{fig:SFH}, we present the star formation histories of the simulated UFDs. In these UFDs, star formation only occurs in the early cosmic times at high redshift ($z>4$), leading to old stellar populations with average ages ranging from 12.67\,Gyr to 12.99\,Gyr, as listed in Table~\ref{tab:properties}. Star formation in these galaxies begins at different redshifts, spanning the range from $z=18.6$ to $z=6.4$. In general, star formation in UFDs proceeds in a bursty mode, particularly when the galaxies host fewer than $10^4\,\Msun$ in stars. In this case, the stellar mass can increase by a factor of several over a brief interval of only a few tens of Myr.

The dashed curves represent the histories of in-situ star formation, defined as the stars formed in the most massive progenitors of the merger tree. A general trend is that more massive galaxies contain a larger fraction of in-situ formed stars, whereas in the m4.5 and m3.6 galaxies, the majority of stars formed ex-situ and were incorporated into the galaxy through merger events.

The reionization time of each halo is marked with crosses on Fig.~\ref{fig:SFH}. We will quantify the timescale of reionization quenching later in Section~\ref{sec:quench}. Here, we stress that reionization does not instantaneously quench any of these galaxies. Instead, they continue forming stars for several hundred Myr after $z_\text{reion}$. Nonetheless, none of these UFDs have any star formation after the reionization quenching. We define the quenching time as the cosmic time when 99\% of the member stars in the galaxy are formed. The m3.6 galaxy experiences the earliest quenching, at 1.03\,Gyr after the Big Bang ($z=5.7$), whereas the m5.5 galaxy quenches the latest, at a cosmic time of 1.62\,Gyr ($z=4.3$).

\subsection{The stellar mass--halo mass relation}
\begin{figure}
	\includegraphics[width=\columnwidth]{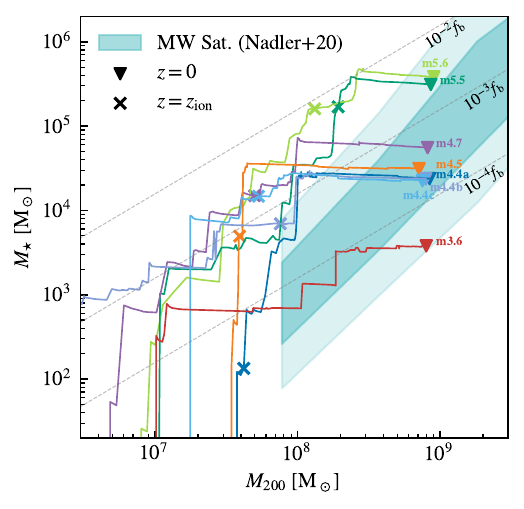}
    \caption{Stellar mass evolution as a function of the halo mass of dwarf galaxies simulated with RIGEL. The triangles mark the final masses at $z=0$, while the crosses mark the masses at $z=z_\text{reion}$. The shaded regions are the stellar-mass--halo-mass relation inferred from Milky Way satellites by \cite{2020ApJ...893...48N:Nadler}.
    }
    \label{fig:SHR}
\end{figure}
Figure~\ref{fig:SHR} shows how the stellar mass $M_\star$ evolves with the halo mass $M_{200}$ of dwarf galaxies. Here, the stellar mass and halo mass refer to the mass of the main progenitor in the merger tree. The stellar mass is computed by summing the masses of all stellar particles within the virial radius ($R_{200}$) of the main halo. The halo properties at $z=0$ are marked with triangles, while the properties at $z=z_\text{reion}$ are marked with crosses.

As our halo selection criterion was controlled, all halos reach a similar halo mass of $\sim10^9\,\Msun$ at $z=0$. However, the final stellar mass exhibits a spread of 2~dex. We find that the final stellar mass has a clear positive correlation with the progenitor halo mass during reionization ($M_\text{halo}^\text{reion}$, see Section~\ref{sec:correlations}). More massive galaxies reside in more massive halos when they are ionized at $z = 5\text{--}7$. Because they assemble earlier and more rapidly, they have already formed more stars and accumulated more self-shielding gas before the radiation can evaporate their surrounding cool CGM and IGM (see Section~\ref{sec:quench} about how reionization quenches UFDs).

Five out of eight (62.5\%) simulated galaxies fall within the 16--84\% confidence interval of the stellar mass--halo mass relation (SHMR) derived from Milky Way (MW) satellites \citep{2020ApJ...893...48N:Nadler}, while the remaining three lie within the broader 5--95\% confidence interval. It is reasonable to compare the simulated isolated galaxies with the MW satellites because the faint-end of the SHMR of MW satellites is primarily shaped by cosmic reionization as well \citep{2000ApJ...542..535G:Gnedin,2014MNRAS.444..503N:Noh,2020MNRAS.498.4887B:Benitez-Llambay}. Thus, this result suggests that our sample is likely a reasonable representation of the observed UFD population.

The stellar mass grows in a stepwise fashion and then slowly declines as mass is lost through stellar feedback. This indicates the bursty nature of star formation associated with the halo merger event. The first starburst typically happens with a halo mass of $\sim10^7\,\Msun$, which is a critical halo mass for galaxy formation via molecular hydrogen cooling \citep{2025ApJ...983L..23N:Nadler}. 

None of the halos shows a substantial increase in stellar mass once its halo mass exceeds $3\times10^8\,\Msun$. This suggests that these halos acquire most of their mass through accretion and dry mergers involving only dark matter. This underscores how crucial the early halo growth histories of UFDs are. Moreover, this implies that later halo growth has a strong impact on the stellar dynamics and on the signatures left by star formation and feedback, since the halo must still accumulate a substantial amount of mass after star formation has been quenched.

\subsection{Luminosity-size relation and Dynamics}
\begin{figure}
	\includegraphics[width=\columnwidth]{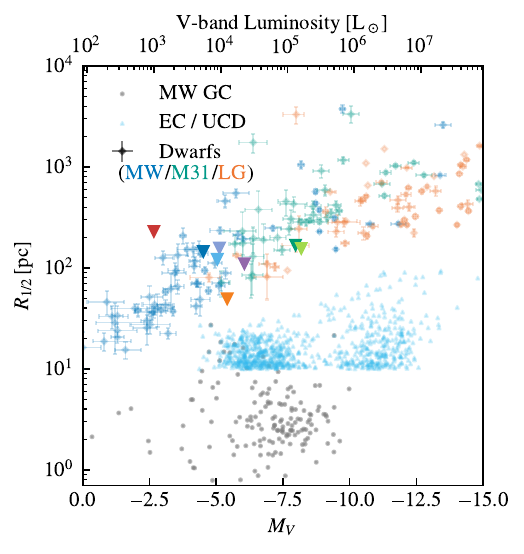}
    \caption{Luminosity--size relation of the simulated UFDs, shown with the same color scheme as Fig.~\ref{fig:SHR}. The simulation results are compared with the observational data of local dwarf galaxies \citep[errorbars;][]{Pace2025OJAp....8E.142P}, MW GCs \citep[grey dots;][]{1996AJ....112.1487H:Harris}, and ECs or UCDs in galaxy clusters  \citep[blue dots;][]{2012A&A...547A..65B:Bruns}.
    }
    \label{fig:MSR}
\end{figure}
In Fig.~\ref{fig:MSR}, we show the relation between the V-band absolute magnitude $M_V$ and the projected
half-light radius $R_{1/2}$ of the simulated UFDs. We compare our results with observational data of nearby dwarf galaxies from the compilations of \cite{Pace2025OJAp....8E.142P}. Moreover, we also present the observations of MW globular clusters \citep[GCs][]{1996AJ....112.1487H:Harris} and extended clusters (ECs) or ultra-compact-dwarf galaxies \citep[UCDs,][]{2012A&A...547A..65B:Bruns}.
To obtain the photometric properties of the simulated galaxies, we calculate the dust-free luminosity of each star particle using the method outlined by \cite{2020MNRAS.492.5167V:Vogelsberger}. The star particles are considered single-age star populations (SSP) without any extinction. This is reasonable, as the UFDs are all old stellar systems without any ISM component. The typical absolute magnitude of an individual star particle is $M_V \approx 2.9$, which corresponds to an apparent magnitude of $m_V = 22.9$ at a typical Milky Way UFD distance of $100$\,kpc \citep[e.g.][]{2012AJ....144....4M:McConnachie}. As this apparent magnitude lies just above the detection threshold of a DES-like survey, we treat the total luminosity of all star particles as the luminosity of the central galaxy. We emphasize that, for naming the galaxies and the stellar mass--halo mass relation, we include all stars within the halo, whereas in the scaling relations presented in this section, we omit stars that reside in satellite systems.

Nearly all dwarf galaxies fall within the observational scatter of the $M_V$--$R_{1/2}$ relation. The smallest system, m3.6, is more diffuse than observed galaxies of similar mass. This behavior is common for simulated dwarf galaxies with $<10^4\,\Msun$, as they may have a substantial fraction of faint stars that are invisible in observations \citep[e.g][]{2025ApJ...978..129A:Andersson,2025ApJ...986..214G,2025ApJ...995..162W:Wheeler}. As this issue has been extensively discussed in recent studies, we only present the intrinsic half-light radius (without detection limits) in this paper. The diffuse intrinsic shape of m3.6 is due to its assembly by several stellar systems with comparable masses, each containing $\lesssim10^3\,\Msun$ stars. The low-mass stellar system is vulnerable to tidal forces during the merger processes. Consequently, its stellar component has become expanded due to repeated and intense dynamical interactions. In contrast, the two most massive galaxies exhibit very similar masses and sizes, as they quickly assemble a massive stellar system with high binding energy.

As Fig.~\ref{fig:Vband} shows, all the galaxies present an elliptical morphology, and their velocity dispersion ranges from $1.28$\,\kms to $4.90$\,\kms. To further examine the stellar kinematics, we calculate the rotation velocity $v_\text{rot}$ and the ellipticities $e$ of our simulated UFDs. Traditionally, the observed $v_
\text{rot}/\sigma_\star$ quantity is the maximum rotational velocity over the central velocity dispersion of a galaxy. To mimic this, we define $v_\text{rot}$ as the maximum value of the stellar rotation curve, and $\sigma_\star$ as the standard deviation of the stellar velocities along the $z$-axis, measured within the radius at which the rotation curve reaches its peak.

To calculate the ellipticities of our simulated galaxies, we use the following method: (\textit{i}). We find the light center of each galaxy based on their V-band luminosity. When calculating the center of light, we only consider the star particles within 3~kpc from the center and determine the final center iteratively; (\textit{ii}). For the intrinsic 3D ellipticity, we compute the mass-weighted covariance matrix of star-particle coordinates relative to the light center. The eigenvalues of this matrix are the intrinsic RMS lengths of the three principal axes $(a,b,c)$. Conventionally, $a\geq b\geq c$ and the 3D ellipticity is $e_{\rm 3D}=1-c/a$; (\textit{iii}). For the projected 2D ellipticity, we project the same star particles along 30 viewing directions uniformly sampled on the sphere. For each projection, we construct a V-band luminosity-weighted image and compute the luminosity-weighted covariance matrix of the image. The eigenvalues of this matrix give the projected RMS lengths of the two principal axes $(a,b)$, with $a\geq b$ and $e_{\rm 2D}=1-b/a$. We report the 16th, 50th, and 84th percentiles of $e_{\rm 2D}$ over the 30 viewing directions.

\begin{figure}
	\includegraphics[width=\columnwidth]{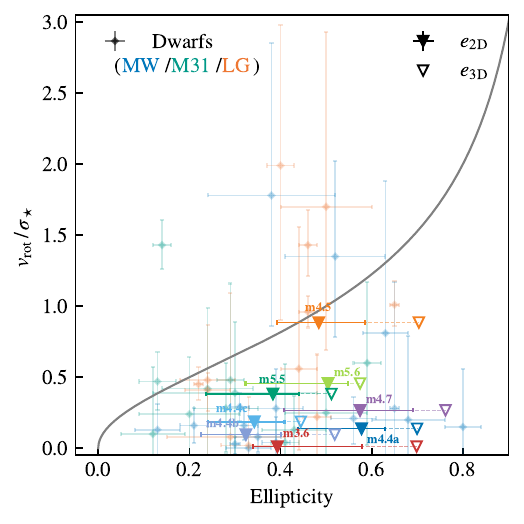}
    \caption{Stellar kinematics of our simulated UFDs: rotation support $v_\text{rot}/\sigma_\star$ versus ellipticity $e$. The errorbars are the observations of nearby dwarf galaxies compiled by \cite{2017MNRAS.465.2420W}. The grey curve represents the predicted behavior of self-gravitating systems that are flattened purely by rotation \citep{1978MNRAS.183..501B:Binney}. Systems below this curve are expected to be dispersion-supported.
    }
    \label{fig:dyn}
\end{figure}

In Fig.~\ref{fig:dyn}, we show the $v_\text{rot}/\sigma_\star$--$e$ relation for our simulated galaxies, and compare it with measurements for nearby dwarf galaxies compiled by \cite{2017MNRAS.465.2420W}, as well as with the expected trend for an idealized, purely rotation-supported system \citep[grey curve,][]{1978MNRAS.183..501B:Binney}. All of our UFDs are located in the dispersion-supported area. Compared with the observations, our simulated UFDs occupy the same region as the observed MW UFDs in this plane. 

Notably, the simulated galaxies show a tendency toward higher ellipticities than the observed local dwarf population. The median projected ellipticities of our galaxies range from $e_{\rm 2D}=0.32$ to $0.58$ (Table~\ref{tab:properties}), with six of the eight galaxies having median values above the observed median of $e=0.365$. The somewhat higher projected ellipticities of our sample may be partly related to the isolated nature of the simulated UFDs. For satellites of the MW and M31, tidal interactions with their host galaxies can modify their stellar morphology through tidal stripping and dynamical heating \citep[e.g.][]{2001ApJ...547L.123M,2015MNRAS.447.1112B:Barber,2018MNRAS.476.3816F:Fattahi}, whereas such environmental evolution is absent in our isolated systems. 

In addition, projection effects can substantially lower the observed ellipticity of our simulated galaxies. As shown in Table~\ref{tab:properties}, their intrinsic three-dimensional ellipticities range from $e_{\rm 3D}=0.44$ to $0.76$, indicating that they are highly elongated elliptical systems. The median reduction due to projection is $\Delta e\simeq0.16$ across our sample. This motivates us to further investigate the physical origin of these intrinsically high ellipticities of the isolated UFDs.

We noticed that \cite{2024MNRAS.527.2403G:Goater} reported that UFDs located in more slowly assembling halos generally show higher ellipticities. They interpret this trend as arising from the reduced in-situ and enhanced ex-situ stellar fractions in such slowly growing halos. The angular momentum brought in by ex-situ stars can indeed account for the high ellipticity of the 4.5 galaxy, which obtains roughly two-thirds of its stellar mass through ex-situ mergers. However, in our sample, even the m5.6 and m5.5 galaxies, which assemble the quickest and form nearly all their stars in situ, nonetheless display ellipticities of $e>0.5$. Although these halos increase in mass through mergers with starless halos, such interactions are unlikely to transfer angular momentum from the DM to the stellar components, given the typical size of our simulated UFDs ($\sim100$\,pc). This implies that these galaxies should be elliptical as they are born. Except for m4.5, all the other galaxies present even higher ellipticities at the time when their star formation is quenched ($z_\text{quench}$)\footnote{Given the typical $\sim30$ Myr crossing time of UFDs, the system can attain dynamical equilibrium shortly after the merger induced starburst via violent relaxation \citep{1967MNRAS.136..101L}. Thus, it is reasonable to evaluate the ellipticity at the quenching time.}.

We explain the high ellipticities of UFDs by the bursty star formation at high redshift and the reionization quenching of the UFDs. At high redshift, the UFDs form most of their stars in intense starbursts triggered by the frequent mergers of gas-rich halos. The gas accreted through these mergers typically carries significant and often misaligned angular momentum, and newly formed stars inherit this kinematic structure, resulting in anisotropic and elongated stellar configurations. Once star formation is quenched by the external UVB, these systems effectively cease further dynamical evolution due to the long relaxation times. Assuming the typical mass-size relation of observed dwarf galaxies \citep[$R\propto M^{0.3}$, e.g.][]{2016MNRAS.462.1470L:Lange,2021ApJ...922..267C:Carlsten}, the relaxation timescale ($t_\text{relax}\propto M^{1/2}R^{3/2}/\ln M$) will be longer than 13.6\,Gyr for systems more massive than $\sim10^4\,\Msun$. Although the m3.6 galaxy is less massive than this critical mass, its large size results in a correspondingly long relaxation time. Therefore, stellar distributions preserve the elongated shapes formed during their initial assembly, and any mergers with other stellar systems tended to enhance this elongation. Consequently, the UFDs exhibit consistently high ellipticities at $z=0$.

\subsection{Metallicities}
\label{sec:MZR}
\begin{figure}
	\includegraphics[width=\columnwidth]{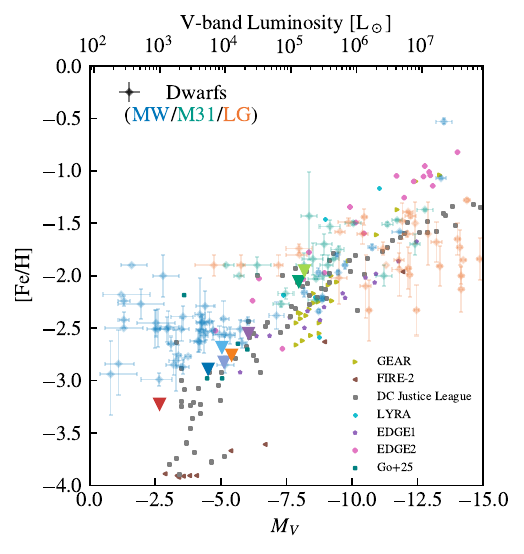}
    \caption{Luminosity-metallicity relation of the simulated UFDs, shown with the same color scheme as Fig.~\ref{fig:dyn}. The simulation results are compared with the observational data of local dwarf galaxies from \cite{Pace2025OJAp....8E.142P}. We notice that the metallicity of these MW UFDs is obtained by narrowband CaHK imaging while the metallicity from \cite{2012AJ....144....4M:McConnachie} is obtained through medium resolution spectroscopy by either spectral synthesis or calcium triplet (CaT) calibration. The colored dots are simulated dwarf galaxies from recent cosmological zoom-in simulations: GEAR \citep{2018A&A...616A..96R:Revaz}, FIRE-2 \citep{2019MNRAS.490.4447W:Wheeler}, DC Justice League \citep{2021ApJ...906...96A:Applebaum}, LYRA \citep{2022ApJ...941..120G:Gutcke}, EDGE1 and EDGE2 \citep{2020MNRAS.491.1656A,2025MNRAS.541.1195R:Rey}, and \cite{2025ApJ...986..214G}.
    }
    \label{fig:MZR}
\end{figure}

In Fig.~\ref{fig:MZR}, we present the luminosity-metallicity relation of the simulated UFDs at $z=0$ and the observational data compiled by \cite{Pace2025OJAp....8E.142P}.  For bright dwarf galaxies with $M_V < -8$ ($L_V\gtrsim10^5\Lsun$), a tight and universal relation between luminosity (or mass) and stellar metallicity is observed \citep{2013ApJ...779..102K:Kirby}. 
\cite{2020MNRAS.491.1656A} shows that the mass--metallicity relation (MZR) is a critical indicator of the feedback model, as it is directly
related to the strength of galactic outflows in dwarf galaxies. The two most luminous galaxies reproduce the metallicity of the observed systems, implying that the treatment of radiation, stellar winds, and SN feedback in the multi-phase ISM sets the feedback strength in a self-consistent manner. 

Although the MZR relation can be extended to the UFD regime, the observed dispersion in metallicity at fixed luminosity increases, and a plateau of metallicities appears around [Fe/H]$\sim-2.5$ \citep{2019ARA&A..57..375S:Simon}. However, numerical simulations typically produce metallicities that are lower than those observed, along with a steeper MZR for UFDs \citep[e.g.][]{2017ApJ...848...85J,2018A&A...616A..96R:Revaz,2019MNRAS.490.4447W:Wheeler,2021ApJ...906...96A:Applebaum}, as shown in Fig.~\ref{fig:MZR} with the colored dots. This metallicity plateau observed in low-mass UFDs can be interpreted as the result of overshooting enrichment from Pop III stars \citep{2023A&A...669A..94S:Sanati,2025arXiv251005232R:Rey,Liu2025,Asada2026} and as a signature of a top-heavy IMF among metal-poor stars \citep{2022MNRAS.513.2326P}.

Since we do not include any dedicated physics for Pop III stars, the faintest galaxies ($L_V\lesssim10^4\Lsun$) in our simulation also exhibit relatively low metallicities. Despite this, our faintest galaxies are substantially more metal rich than those found in the FIRE-2 \citep{2019MNRAS.490.4447W:Wheeler} and DC Justice League \citep{2021ApJ...906...96A:Applebaum} simulations. This difference can be attributed to the inclusion of radiative feedback in our RIGEL framework, which leads to a more gentle regulation of star formation with weaker
galactic outflows \citep{2025MNRAS.541.1195R:Rey}. As a result, a longer star formation duration will allow more iron enrichment by Type Ia SN and more metals will be retained in the galaxy. Moreover, \cite{2025ApJ...986..214G} showed that the individual IMF sampling, as used by RIGEL, can also contribute to the higher metallicities found in simulated UFDs because it maintains enrichment as a local and time-resolved process, thereby allowing metals to be locked into stars more efficiently.

\subsection{Dark matter halo profiles}
Lastly, we examine the properties of DM halos hosting these simulated UFDs. The DM halos of dwarf galaxies are closely related to the small-scale challenges to the $\Lambda$CDM paradigm \citep{2017ARA&A..55..343B:Bullock}. Over the past twenty years or so, the cusp--core problem, namely that dark-matter--only (DMO) $\Lambda$CDM simulations produce halos with steep central cusps, $\rho\propto r^{\alpha}$ with $\alpha\leq -1$ \citep{1997ApJ...490..493N:Navarro}, whereas observations of dwarf galaxies seem to favor approximately constant-density cores at small radii \citep{2001AJ....122.2381M:McGaugh}, has been largely resolved through the inclusion of baryonic feedback. Numerical simulations indicate that stellar feedback can produce cored DM profiles by inducing repeated episodes of gas outflow and accretion \citep{2016MNRAS.456.3542T,2020MNRAS.497.2393L:Lazar}. In this feedback framework, UFDs that exhibit a low stellar-mass--halo-mass ratio are assumed to experience minimal feedback, and their halos are therefore expected to remain cuspy. However, recent observations of MW UFDs \citep{2020ApJ...904...45H:Hayashi,2023ApJ...953..185H:Hayashi} have reported systems hosted by cuspy DM halos as well as by cored ones, indicating that the host halos of UFDs exhibit structural diversity.

\begin{figure*}
	\includegraphics[width=2\columnwidth]{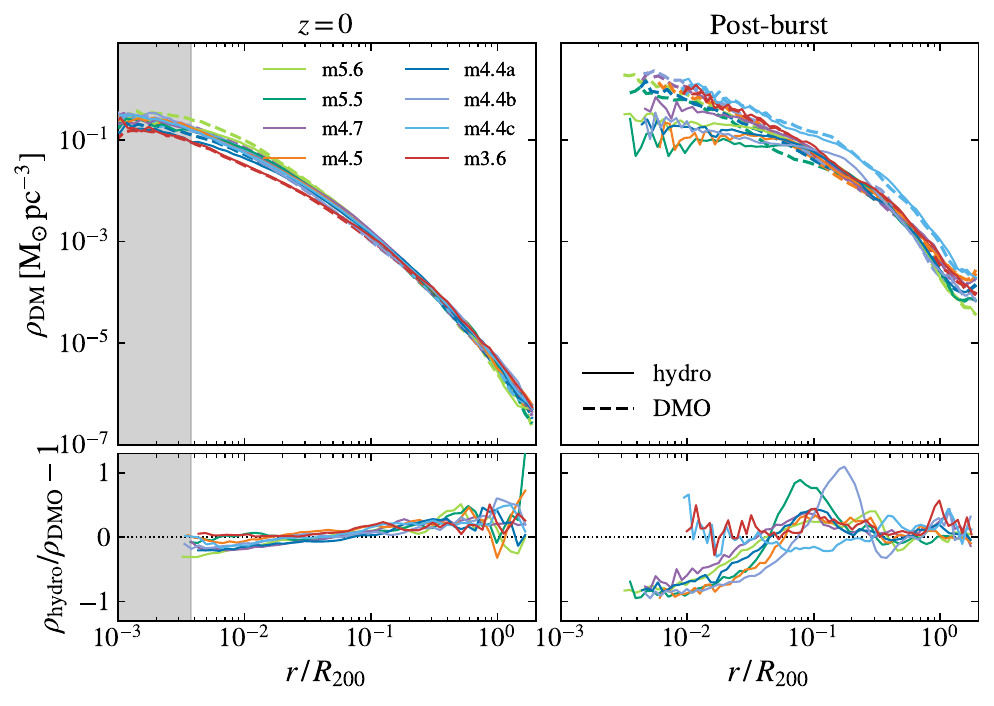}
    \caption{Dark matter density profiles of the eight RIGEL zoom-in halos at $z=0$. The left column shows $z=0$; the right column shows the post-starburst epoch. Top panels show $\rho_{\rm DM}$ as a function of radius normalized by $R_{200}$. Solid lines are the RIGEL runs; dashed lines of the same color are the corresponding DMO counterparts, scaled by $1-\Omega_{\rm b}/\Omega_{\rm m}$. Bottom panels show the relative difference $\rho_{\rm RIGEL}/\rho_{\rm DMO}-1$. Inner bins below each halo's own resolution radius are omitted from the residual panel. The grey band in the left column indicates the unresolved inner region, which is defined as the largest radius among all RIGEL and DMO halos that still contains fewer than $3000$ DM particles. This radius is generally comparable to the gravitational softening length of the DM particles ($\epsilon_\text{DM}=69$\,pc). We do not show the resolution limit because $R_{200}$ varies substantially between different halos at high redshift.}
    \label{fig:DM_profiles}
\end{figure*}

In Fig.~\ref{fig:DM_profiles}, we demonstrate the DM density profiles of the simulated UFDs and compare them with the corresponding profiles in DMO simulations. For all the galaxies, the RIGEL and DMO simulations have minor differences in their profiles. This is because there is no star formation activity during most of the halo growth history. As shown in Fig.~\ref{fig:SHR}, even the most rapidly growing halo acquires more than $70\%$ of its mass after the star formation is quenched. Any imprints produced by stellar feedback at high redshift will be readily wiped out by later dark matter accretion and dry mergers \citep{2004MNRAS.349.1117B,2021MNRAS.504.3509O}.

\begin{figure}
	\includegraphics[width=\columnwidth]{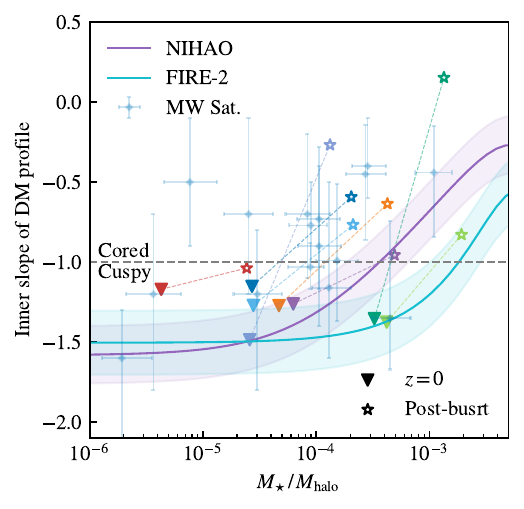}
    \caption{The inner dark matter density slope at 1.5\% of the virial radius as a function of the ratio between stellar and dark-halo masses, shown with the same color scheme as Fig.~\ref{fig:DM_profiles}. The triangles are our simulated UFDs at $z=0$, while the crosses are the observations of MW satellites from \cite{2020ApJ...904...45H:Hayashi} and \cite{2023ApJ...953..185H:Hayashi}. The hollowed stars are the post-burst slope of the UFDs, obtained by taking the shallowest slope found within 100 Myr after the strongest starburst at high-redshift.  The purple and blue shaded regions are the simulated dwarf galaxies from NIHAO \citep{2016MNRAS.456.3542T} and FIRE-2 \citep{2020MNRAS.497.2393L:Lazar} simulations, respectively.
    }
    \label{fig:DM_alpha}
\end{figure}
The central DM profiles exhibit a flattened feature, which is considered a numerical artifact due to the insufficient mass resolution of DM. \citep{2003MNRAS.338...14P:Power} suggested that a minimum of 3000 enclosed DM particles is required in order to achieve converged DM density profiles. We mark this limit with the shaded regions in each panel of Fig.~\ref{fig:DM_profiles}. Despite this artificial structure, none of the eight galaxies show a clearly defined core. The bottom panel of Fig.~\ref{fig:DM_profiles} shows the relative difference between the RIGEL and DMO runs $\rho_{\rm RIGEL}/\rho_{\rm DMO}-1$. The RIGEL and DMO runs exhibit remarkably similar density profiles within $0.1R_{200}$, with no clear systematic offset and relative differences typically below $\sim20\%$. Beyond $0.1R_{200}$, the RIGEL runs show a modest tendency toward higher densities, while the differences become increasingly scattered near $R_{200}$.

To further quantify the non-detection of DM core, we fit the slope of the profiles over the commonly used interval from 1\% to 2\% of the virial radius as the slope evaluated at 1.5\% of $R_\text{vir}$. In Fig.~\ref{fig:DM_alpha}, we present the inner dark-matter density slopes as a function of the stellar-to-halo mass ratio, and we compare our UFD results with the NIHAO \citep{2016MNRAS.456.3542T} and FIRE-2 \citep{2020MNRAS.497.2393L:Lazar} simulations, as well as with measurements for Milky Way satellites from \cite{2020ApJ...904...45H:Hayashi} and \cite{2023ApJ...953..185H:Hayashi}. 

Taking $\alpha = -1$ as a boundary for cusps and cores, all of our UFDs present cuspy DM profiles. This can also be attributed to the brief and prematurely quenched star formation history of UFDs. The m4.4a galaxy presents the shallowest slope of $-1.15$. Five of the galaxies lie in the range predicted by NIHAO or FIRE simulations, while the three relatively small galaxies, m4.4a, m4.4c, and m3.6, present shallower slopes. However, as shown in the left column of Fig.~\ref{fig:DM_profiles}, the profiles including feedback differ only negligibly from the DMO runs, indicating that the variation in slope is unlikely to be caused by baryonic feedback.

In fact, the development of a core in $10^9\,\Msun$ halos demands sustained star formation over nearly the age of the Universe \citep[14\,Gyr;][]{2016MNRAS.459.2573R}. However, at high redshift, these galaxies have much smaller halo masses but very bursty star formation. Consequently, it is much easier to form cores. To examine this, we also fit the high-redshift DM density profiles right after the strongest starbursts. Since DM halos respond to starburst feedback on different timescales, we take the shallowest slope found within 100\,Myr after the strongest starburst and use it as the post-burst slope. In the right panel of Fig.~\ref{fig:DM_profiles}, we show the DM profiles after the burst compared with the DMO runs, and in Fig.~\ref{fig:DM_alpha}, we present the post-burst slopes using hollow star symbols. These galaxies exhibit higher stellar-to-halo mass ratios and flatter DM profiles. The majority of them (six out of eight) display a shallower slope than that found in the NIHAO or FIRE simulations. Except for the m3.6 and m4.4c galaxies, every other galaxy develops a post-starburst DM profile that is significantly shallower than in the DMO case, with $\alpha > -1$. The m4.4b and m5.5 galaxies present prominent cores with almost flat inner DM profiles ($\alpha\approx0)$. These findings indicate that baryonic feedback can indeed shape the DM halo on relatively short timescales. Nevertheless, for UFDs whose star formation is quenched by reionization, it is difficult to preserve a DM core in the absence of ongoing star formation. Consequently, as the halo grows, the structure of UFD halos is mainly governed by gravitational evolution and tidal interactions, leading to predominantly cuspy DM profiles in present-day UFDs.

\section{Impacts and imprints of inhomogeneous reionization}
\label{sec:reion}
In the preceding section, we demonstrated that the eight simulated UFDs with comparable halo masses exhibit diversity in their stellar masses, kinematics, metallicities, and central DM profiles. Among these different types of diversity, we have shown that the variation in DM profiles arises primarily from the assembly history of halos, whereas the other forms of diversity are closely linked to inhomogeneous reionization. In this section, we explore how reionization affects UFDs and examine the imprint left by its inhomogeneous nature.

\subsection{Reionization-driven diffuse gas removal dominates quenching}
\label{sec:quench}

\begin{figure*}
	\includegraphics[width=2\columnwidth]{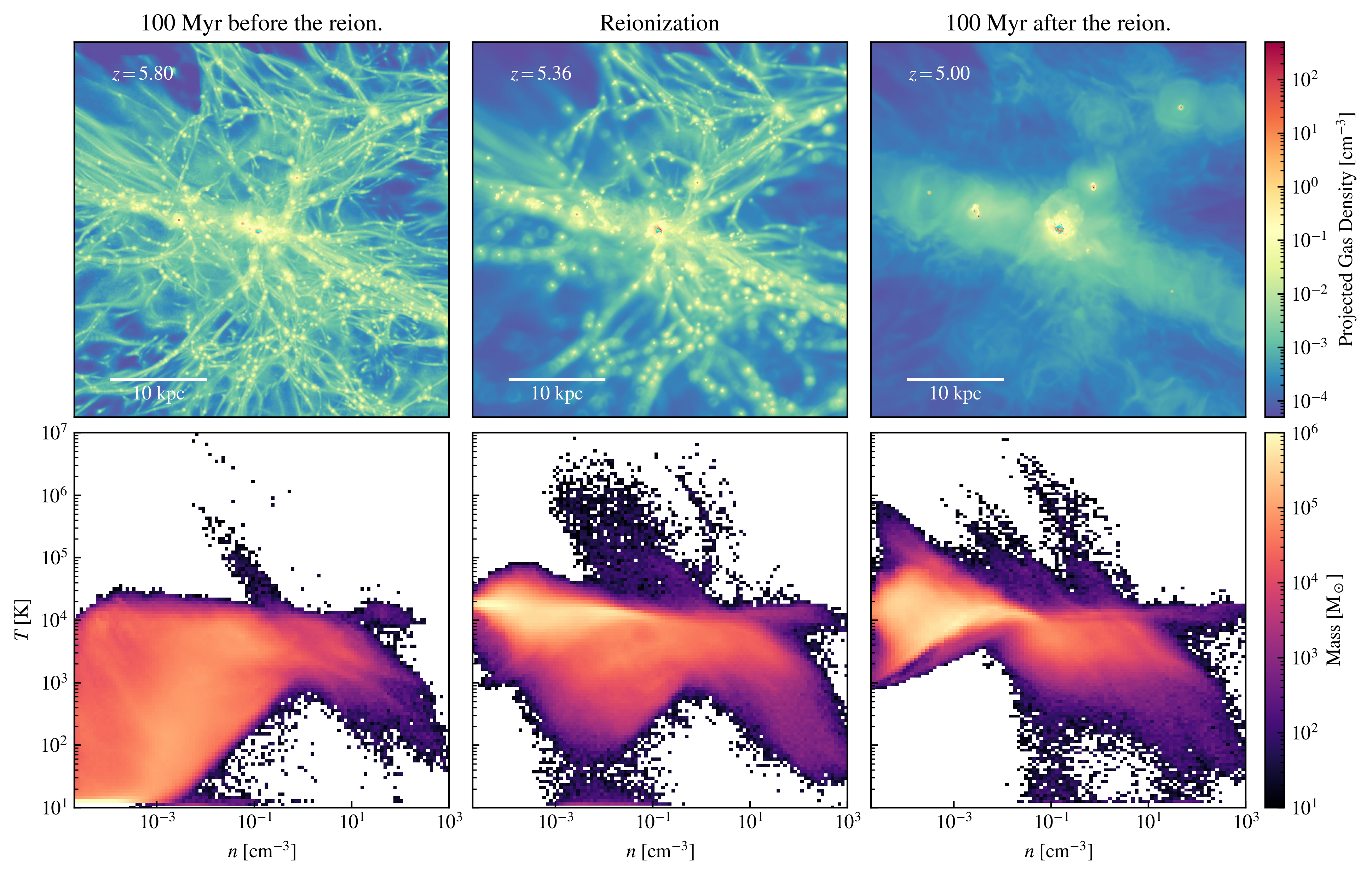}
    \caption{Density-weighted projections of the gas number density (top panels) and phase diagrams (bottom panels) of the m5.5 galaxy. Panels show snapshots captured at 100\,Myr before reionization (left), during reionization (middle), and after reionization (right).
    }
    \label{fig:projection}
\end{figure*}
In this section, we examine how reionization quenches star formation in our simulated UFDs. \cite{2024MNRAS.528.1296C:Chan} separated the progression of the reionization process into three phases. First, during the propagation phase, the I-front sweeps across the halo at a small fraction of the speed of light, ionizing and heating the low-density IGM. Next, in the photoevaporation phase, the I-front propagates into halos, ionizing the gas in their outer regions. In the final relaxation phase, the thermodynamic structure of the halos relaxes and settles into a new equilibrium.

Focusing on the suppression of star formation within the host halo, we find that this phase is more accurately described as a two-stage process. During both the I-front propagation and photoevaporation phases, reionization primarily acts to photoionize and heat the diffuse neutral gas around the galaxy, thereby depleting the reservoir that would otherwise resupply the ISM. Star formation then continues for a finite period using the surviving self-shielded gas, but eventually ceases once this gas is consumed or dispersed by stellar feedback and can no longer cool and recollapse in the presence of the UV background.
In Fig.~\ref{fig:projection}, we illustrate the impact of reionization on the gas surrounding the m5.5 galaxy. We show projected gas density maps within a $(40\,{\rm kpc})^3$ box centered on the galaxy together with the corresponding gas phase diagrams. The three columns correspond to snapshots taken approximately $100$\,Myr before reionization, during the passage of the I-front, and $100$\,Myr after reionization. The box size roughly corresponds to the Lagrangian volume of dark matter particles that will eventually reside within the halo virial radius at $z=0$.

Prior to reionization, the IGM exhibits a filamentary and clumpy structure (top-left panel). In the absence of photoheating, most diffuse gas remains neutral and cold, resulting in low thermal pressure. As shown in the bottom-left panel, gas with densities $n<10^{-1}\,{\rm cm^{-3}}$ occupies a broad temperature range between $\sim10$\,K and $10^4$\,K. Because the cooling time is shorter than the halo dynamical time, gas can continuously condense from the IGM onto the galaxy, replenishing the ISM and sustaining a self-regulated cycle of star formation and stellar feedback. Although mergers can temporarily disrupt this equilibrium, continued accretion from the surrounding IGM rapidly restores the gas supply.

The situation changes dramatically when the I-front arrives. Gas exposed to the UV background is heated to temperatures of $\sim1.5\times10^4$\,K, while only gas above the self-shielding threshold, approximately $n_{\rm ssh}\simeq4\times10^{-3}\,{\rm cm^{-3}}$ \citep{2013MNRAS.430.2427R}, remains largely neutral. The timescale required to photoheat the entire Lagrangian volume can be approximated by the I-front crossing time, $t_{\rm ic}\approx20\,{\rm kpc}/1000\,{\rm km\,s^{-1}}\approx20\,{\rm Myr}$, which is much shorter than the halo dynamical time. Consequently, as shown in the middle panels of Fig.~\ref{fig:projection}, the diffuse cold and cool gas is rapidly ionized and heated. The filamentary structures visible before reionization disappear almost entirely, leaving only dense gas associated with minihalos and the ISM of the central galaxy. We note, however, that this transition may be less abrupt when accounting for additional pre-reionization processes, such as X-ray pre-heating and baryon–dark matter streaming velocities, both of which can suppress small-scale gas clumping prior to the arrival of the I-front \citep{2021ApJ...908...96P:Park}.

\begin{figure*}
	\includegraphics[width=2\columnwidth]{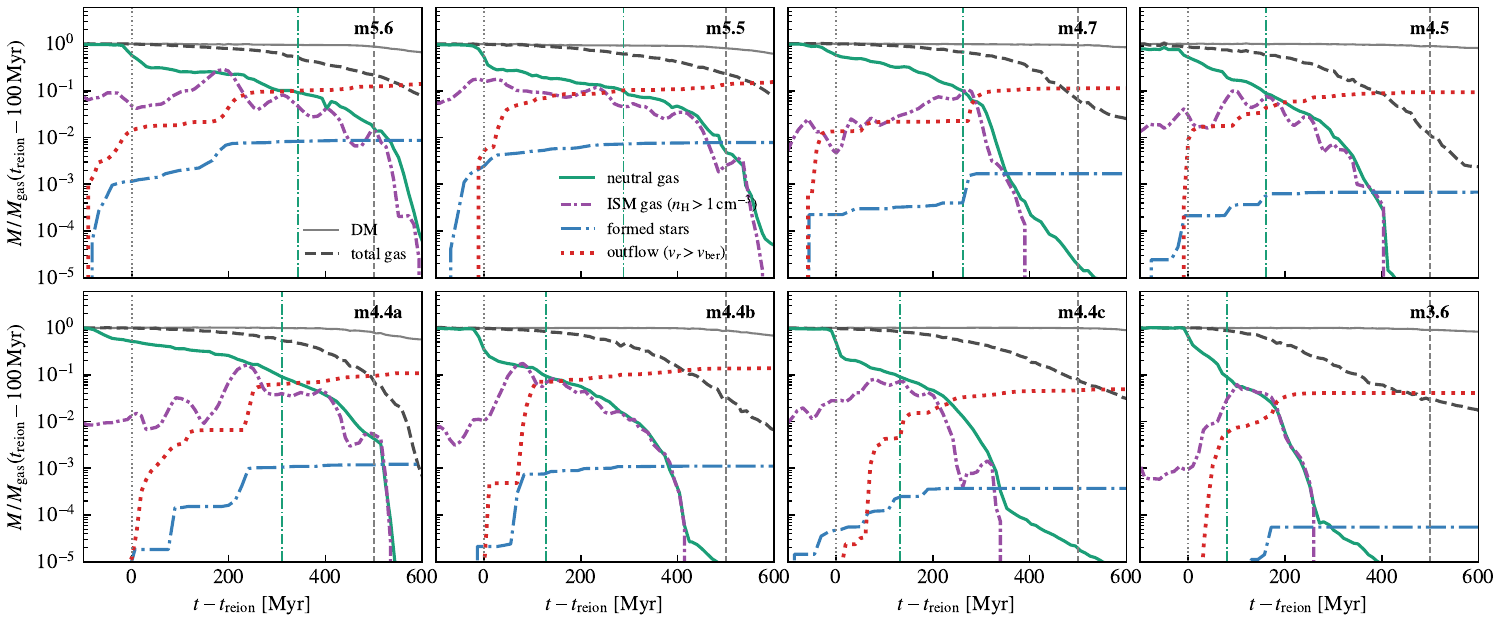}
    \caption{Evolution of the DM, gas, stellar, and outflow mass components within the Lagrangian volumes of the eight simulated ultra-faint dwarf galaxies. The x-axis shows the time relative to the local reionization epoch, $t_\text{reion}$, while all masses, except the DM mass, are normalized by the total gas mass within the Lagrangian volume at $t_\text{reion}-100\,\text{Myr}$. The grey curve shows the dark matter mass enclosed within the Lagrangian volume normalized by the DM mass at $t_\text{reion}-100\,\text{Myr}$. Black, green, and purple curves show the total gas mass, neutral gas mass, and ISM gas mass ($n_{\rm H}>1\,{\rm cm^{-3}}$), respectively. The blue curve shows the cumulative mass of stars formed after the first tracked snapshot, and the red curve shows the cumulative mass removed by hot outflows with $v_r>v_{\rm ber}$. The vertical dotted line indicates the epoch of reionization, and the vertical dashed line marks the point $t_\text{reion}+500\,\text{Myr}$ used for the gas budget statistic in Fig.~\ref{fig:channel_frac}. The green dash-dotted line indicates the moment when the neutral gas reaches 10\% of its mass at $t_\text{reion}-100\,\text{Myr}$. The relative evolution of these components illustrates the depletion of cold gas, the decline of star formation, and the varying importance of gas retention and outflows among different UFDs following reionization.
    }
    \label{fig:mass-budget}
\end{figure*}

The thermodynamic transformation of the diffuse gas does not immediately remove it from the Lagrangian volume. This can be seen more clearly in Fig.~\ref{fig:mass-budget}, which illustrates the time evolution of the gas content for all eight UFDs. We define the Lagrangian volume by tracing back the dark matter particles that reside within the virial radius of the UFD halo at $t_\text{reion}+600\,\text{Myr}$. The total gas budget is then obtained by summing the gas mass contained in this Lagrangian volume at $t_\text{reion}-100\,\text{Myr}$. As indicated by the grey solid curves in Fig.~\ref{fig:mass-budget}, the total dark matter mass within the Lagrangian region remains nearly constant, whereas by $t=t_\text{reion}+600\,\text{Myr}$ at least 90\% of the gas is removed. This demonstrates that changes in the Lagrangian volume have only a minor impact on the gas content. During the first $\sim100$\,Myr after reionization, the total gas mass within the Lagrangian region remains nearly constant. In contrast, the neutral gas mass (\HI and \ce{H2}) declines rapidly following the arrival of the I-front. At the same time, the ISM gas mass, defined as gas with $n_{\rm H}>1\,{\rm cm^{-3}}$, remains approximately unchanged and continues to account for roughly $1$--$10\%$ of the total gas mass.

Reionization therefore first destroys the diffuse neutral reservoir rather than directly removing the dense star-forming gas. Star formation can continue temporarily because the cooling time $t_\text{cool}$ of neutral gas is shorter than the photoevaporation timescale in halos with masses $\gtrsim5\times10^6\,\Msun$ \citep{2020ApJ...905..151N}. Even the smallest m3.6 galaxy meets this criterion, so diffuse neutral gas can condense into the ISM faster than it is removed by photoevaporation.

The timescale for this reservoir depletion is naturally associated with photoevaporation. The characteristic sound-crossing time of a minihalo is $t_{\rm sc}=100\,M_7^{1/3}[(1+z)/10]^{-1}\,{\rm Myr}$ \citep{2004MNRAS.348..753S}, where $M_7=M_\text{halo}/10^7\,\Msun$. Before reionization, gas continuously flows from the IGM into the CGM and subsequently replenishes the ISM. Reionization interrupts this supply by photoionizing and gradually removing the diffuse gas surrounding the galaxy. Meanwhile, the remaining self-shielded gas can continue to fuel star formation. Once these dense clumps are consumed or dispersed by stellar feedback, however, the resulting warm gas can no longer efficiently cool and recollapse under the persistent UV background. Reionization therefore couples photoevaporative removal of the diffuse reservoir with starvation of the star-forming ISM \citep{2015Natur.521..192P:Peng}, allowing star formation to persist for several hundred Myr after the arrival of the I-front before ultimately ceasing.

To distinguish this gradual removal from energetic feedback-driven outflows, we compute outflow rates at $r=R_{\rm h,gas}$, where $R_{\rm h,gas}$ is the radius enclosing half of the halo gas mass. Following \citet{2021MNRAS.508.2979P:Pandya}, we apply a Bernoulli velocity criterion ($v_r>v_{\rm ber}$) and count only gas elements that are expected to reach at least $2R_{\rm vir}$. For the halos considered here, the corresponding escape velocity is approximately $v_{\rm esc}\approx15\,{\rm km\,s^{-1}}$, which excludes gas motions associated with pure photoevaporation, whose characteristic velocity is only the sound speed of the photoheated gas, $c_{\rm s}\sim10\,{\rm km\,s^{-1}}$. As shown in Fig.~\ref{fig:mass-budget}, the cumulative outflow mass typically remains below $\sim10\%$ of the ISM mass during the first $100$--$200$\,Myr after reionization, demonstrating that stellar feedback alone has not yet significantly depleted the galaxy's dense gas supply.

\begin{figure}
	\includegraphics[width=\columnwidth]{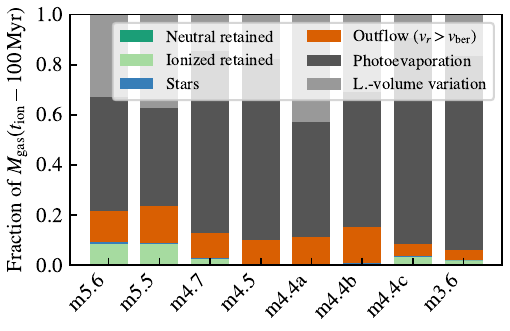}
    \caption{Gas budget of the eight simulated UFDs $500$ Myr after reionization. Each stacked bar shows the fraction of the gas mass present within the Lagrangian volume before the reionization, $M_{\rm gas}(t_\text{reion}-100\,\text{Myr})$, that is found in different channels at $t=t_\text{reion}+500$ Myr. The dark green component shows  neutral gas (\HI+\ce{H2}) retained within the Lagrangian volume, while the light green component shows retained ionized gas. The blue component represents gas converted into stars after reionization, and the orange component denotes gas removed by outflows that satisfy the Bernoulli criterion. The grey component corresponds to the residual mass budget, interpreted as gas lost through photoevaporation and subsequently escape from the Lagrangian volume. The relative contributions of these channels demonstrate that reionization primarily suppresses star formation by removing the diffuse gas reservoir and preventing the replenishment of the ISM, whereas direct removal of gas through stellar-feedback-driven outflows contributes only a minor fraction of the total gas loss.
    }
    \label{fig:channel_frac}
\end{figure}

We quantify the relative importance of these gas-loss channels in Fig.~\ref{fig:channel_frac}. At $t=t_\text{reion}+500,\text{Myr}$, less than $40\%$ of the initial gas mass is retained across all halos, predominantly in a diffuse ionized phase. Because photoevaporative flows cannot be cleanly separated kinematically, we infer the mass lost through this channel as the residual of the gas budget after subtracting gas that remains neutral or ionized, has been converted into stars, or has been removed in energetic outflows. This residual accounts for approximately $45$--$80\%$ of the initial gas mass, whereas energetic stellar-feedback-driven outflows typically account for less than $\sim15\%$, and only a few percent is converted into stars. More massive systems retain up to $\sim30\%$ of their initial gas mass, while the least massive halos retain only a few percent. In nearly all cases, the surviving neutral component constitutes less than $\sim2\%$ of the initial gas reservoir.

Taken together, these results show that reionization quenches the simulated UFDs through a delayed, two-stage process rather than by immediately expelling their star-forming gas. The I-front first ionizes and heats the diffuse neutral reservoir, after which photoevaporation gradually removes this gas and suppresses further replenishment of the ISM. Dense self-shielded gas can therefore sustain star formation temporarily, but once it is consumed or dispersed, the UV background prevents efficient cooling and recollapse. The resulting combination of photoevaporative diffuse-gas removal and starvation, rather than direct ejection by energetic stellar-feedback-driven outflows, sets the quenching of these UFDs.

\subsection{Imprints of reionization on UFD properties}
\label{sec:correlations}

\begin{figure*}
	\includegraphics[width=2\columnwidth]{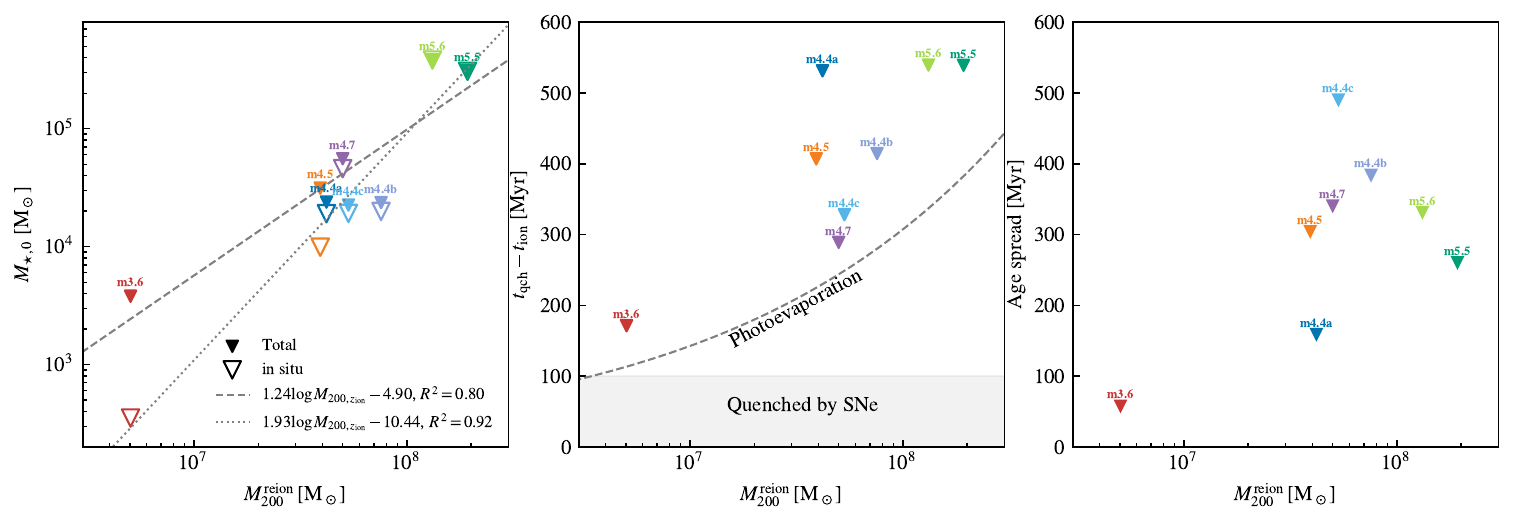}
    \caption{Correlations between UFD properties and $M_{200,z_\text{reion}}$. Left panel: present-day stellar mass $M_{\star,0}$ as a function of $M_{200,z_\text{reion}}$. Solid triangles indicate the total stellar mass, while open triangles denote stars formed in situ within the main progenitor halo. The grey dashed (dotted) line shows a power-law fit to the total (in-situ) stellar mass. Middle panel: time delay between galaxy quenching ($t_\text{qch}$) and halo reionization ($t_\text{reion}$). The quenching time is defined as the epoch by which 99\% of the stars have formed, and the reionization time corresponds to when gas in the shell $2R_\text{vir}<r<3R_\text{vir}$ reaches an ionization fraction of 0.5. The grey shaded region represents the typical timescale over which a UFD is quenched by multiple SNe, while the grey dashed line indicates the photoevaporation timescale of the halo \citep{2004MNRAS.348..753S}. Right panel: stellar age spread as a function of $M_{200,z_\text{reion}}$.
    }
    \label{fig:mion-relations}
\end{figure*}

Having established the mechanism by which reionization quenches star formation, we now examine the long-term imprints that this process leaves on UFDs. Since the formation and evolution of galaxies are governed by the growth of their host DM halo, the evolutionary state of a halo at the moment of reionization plays a key role in shaping its subsequent evolution. In particular, the halo mass at reionization, $M_{{\rm 200},z_\text{reion}}$, sets both the amount of gas available to fuel future star formation and the ability of the halo to retain this gas against photoevaporation. Physically, this mass dependence largely reflects the depth of the halo potential well, which can be characterized by the virial temperature, $T_{\rm vir}\propto M_{\rm vir}^{2/3}(1+z)$, with a weaker dependence on halo structure. At $z\simeq6$, halos spanning $M_{\rm vir}\simeq5\times10^6$--$1.3\times10^8\,\Msun$ have characteristic virial temperatures of approximately $1.5\times10^3$--$1.3\times10^4\,\mathrm{K}$. Since reionization heats ionized gas to a characteristic temperature of $\gtrsim10^4\,\mathrm{K}$, this variation in $T_{\rm vir}$ is critical for determining the susceptibility of a halo to photoevaporation. Gas can escape much more readily from low-mass halos with $T_{\rm vir}\ll10^4\,\mathrm{K}$, whereas the deeper potential wells of halos with $T_{\rm vir}\gtrsim10^4\,\mathrm{K}$ allow them to retain a larger fraction of their photoheated gas. We therefore expect many present-day properties of UFDs to correlate more strongly with $M_{{\rm 200},z_\text{reion}}$ than with either the reionization redshift or the present-day halo mass.

Figure~\ref{fig:mion-relations} summarizes these correlations. In the left panel, we show the relation between $M_{{\rm 200},z_\text{reion}}$ and the present-day stellar mass. We plot both the total stellar mass (solid triangles) and stars formed in situ within the most massive progenitor halo (open triangles). Both the total stellar mass and the in-situ stellar mass exhibit a strong positive correlation with halo mass at reionization. The best-fit power-law relation for the total stellar mass is $\log{M_{\star,0}} =1.24\log{M_{{\rm 200},z_\text{reion}}}-4.90$, while for the in-situ stars, it is $\log{M_{\star,0}} = 1.93\log{M_{{\rm 200},z_\text{reion}}}-10.44$. The slope of the total stellar mass is shallower than that of the in-situ stellar mass. This can be explained by the fact that all these halos share a similar halo mass at $z=0$. Therefore, halos that had a smaller $M_{{\rm 200},z_\text{reion}}$ typically have a lower in-situ mass fraction, as they acquire a larger fraction of ex-situ stars through mergers during their post-reionization growth.

This correlation between $M_{{\rm 200},z_\text{reion}}$ and $M_{\star,0}$ arises because more massive halos can hold on to a larger fraction of their baryons after reionization and sustain star formation for a longer time, as they contain more self-shielded gas prior to reionization. Consequently, even though the eight simulated systems converge to similar halo masses by $z=0$, differences in their masses at the time of reionization produce nearly two orders of magnitude variation in stellar mass.

One might expect that the present-day stellar mass also depends strongly on the stellar mass already assembled when reionization occurs, $M_{\star,z_\text{reion}}$. The m4.4a, m4.4b, and m4.4c galaxies provide a useful test of this possibility, as they have nearly identical present-day stellar masses, $M_{\star,0}\simeq(2.3$--$2.4)\times10^4\,\Msun$, despite having markedly different stellar masses at reionization. The cleanest comparison is between m4.4a and m4.4c, which are reionized at almost the same redshift ($z_\text{reion}=6.15$ and $6.09$, respectively) and have comparable halo masses at that time ($M_{200,z_\text{reion}}=4.2\times10^7$ and $5.3\times10^7\,\Msun$). Nevertheless, their stellar masses at reionization differ by more than two orders of magnitude, from $M_{\star,z_\text{reion}}\simeq1.4\times10^2\,\Msun$ for m4.4a to $1.47\times10^4\,\Msun$ for m4.4c. By $z=0$, however, their stellar masses have converged to $2.39\times10^4$ and $2.25\times10^4\,\Msun$, respectively. Thus, while m4.4c had already assembled $\sim65\%$ of its present-day stellar mass by reionization, m4.4a formed nearly all of its stars afterward.

This comparison demonstrates that $M_{\star,z_\text{reion}}$ alone does not determine the final stellar mass of a UFD. Instead, the substantial post-reionization growth of m4.4a suggests that the duration for which a galaxy can continue forming stars after reionization is also critical. We therefore next examine the delay between reionization and quenching and how this timescale depends on the halo mass at reionization.

The middle panel of Fig.~\ref{fig:mion-relations} shows the delay between reionization and galaxy quenching. All simulated galaxies continue forming stars for at least $\sim150$\,Myr after the arrival of the I-front, even for the smallest galaxy, which is longer than the typical timescale for quenching by multiple SNe. Moreover, reionization quenching takes longer than the timescale of photoevaporation. This result is consistent with the starvation picture described in Section~\ref{sec:quench}. More importantly, the quenching delay increases systematically with $M_{{\rm 200},z_\text{reion}}$. Larger halos retain more self-shielded gas and therefore require longer times to exhaust their remaining fuel supply. The correlation demonstrates that reionization does not quench all UFDs simultaneously; rather, the response depends strongly on the depth of the gravitational potential well at the time of reionization.

The prolonged star formation in more massive systems also leaves a measurable fossil record in their stellar populations. This is illustrated in the right panel of Fig.~\ref{fig:mion-relations}, which shows the stellar age spread (difference between the 84 and 16th percentiles of the stellar age distribution) as a function of $M_{{\rm 200},z_\text{reion}}$. In general, galaxies with larger halo masses at reionization exhibit broader age distributions, reflecting their more extended star formation histories. Longer-lived star formation also allows a larger contribution from delayed enrichment sources such as Type Ia SNe. Consequently, variations in $M_{{\rm 200},z_\text{reion}}$ are expected to produce not only differences in stellar mass but also systematic differences in chemical abundance patterns.

Taken together, these correlations suggest that halo mass at reionization is one of the primary quantities governing the diversity of UFDs. Inhomogeneous reionization introduces variations in both the timing of reionization and the growth histories of halos prior to reionization. As a result, different halos are effectively frozen at different stages of their evolution when the I-front arrives. This variation in $M_{{\rm 200},z_\text{reion}}$ subsequently propagates into differences in stellar mass, quenching timescale, age distribution, and chemical enrichment, thereby generating much of the diversity observed among UFDs at the present day.

\begin{figure}
	\includegraphics[width=\columnwidth]{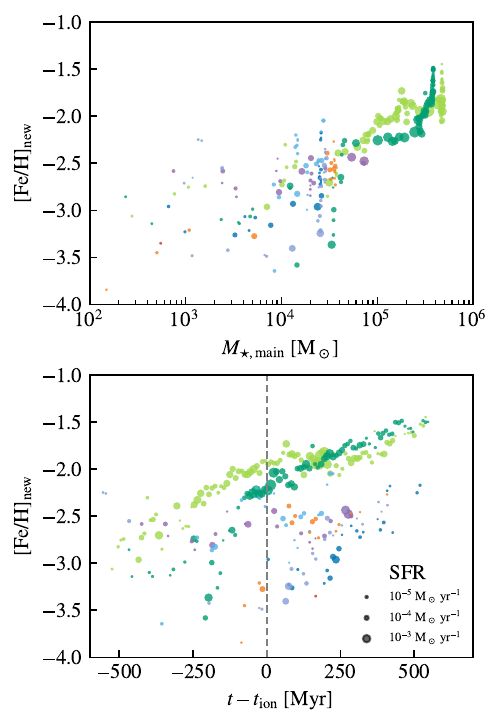}
    \caption{Evolution of the metallicity of newly formed in-situ stars in the eight simulated UFDs. Top panel: mass-weighted metallicity of stars formed within each output interval as a function of the total stellar mass of the main halo. Bottom panel: the same quantity plotted against time relative to reionization. Marker size scales with the star formation rate in that interval.}
    \label{fig:FeH_reionization}
\end{figure}

The prolonged star formation in more massive systems also leaves a measurable imprint on their chemical enrichment histories. Figure~\ref{fig:FeH_reionization} shows the metallicity of newly formed in-situ stars as a function of both the stellar mass of the main progenitor and the time relative to reionization. Each point represents the stars formed within a single time interval between two snapshots ($\sim10$\,Myr), with the symbol size indicating the stellar mass formed during that interval. 

The top panel of Fig.~\ref{fig:FeH_reionization} shows that the metallicities of newly formed stars present a large scatter when the stellar mass of the galaxy is smaller than $\sim 3\times10^4\,\Msun$. This indicates that the early star formation in these UFDs is generally triggered by gas inflow or gas-rich mergers, leading to a bursty pattern. On the contrary, the metallicity increases smoothly along with galaxy growth at higher stellar masses. Interestingly, in both the m5.6 and m5.5 galaxies (green dots), we found a rapid rise in iron abundance with only a modest growth in total stellar mass, immediately before star formation is quenched. This is a clear signature of Type Ia SNe.

Prior to reionization, newly formed stars typically exhibit metallicities of ${\rm [Fe/H]}\lesssim-2.5$, with substantial scatter among different halos. This diversity reflects differences in the assembly histories and early enrichment of the progenitor systems. After reionization suppresses cosmological gas accretion, star formation can continue for several hundred Myr using the gas already retained within the halos. Continued enrichment of this remaining gas therefore leads to a progressive increase in the metallicity of newly formed stars, reaching ${\rm [Fe/H]}\sim-1.5$ in several systems, significantly higher than that of the pre-reionization stellar populations.

This behavior is a direct consequence of the delayed quenching discussed above. Reionization rapidly suppresses the inflow of fresh gas, but star formation continues for several hundred Myr using the surviving self-shielded gas reservoir. During this period, metals released by SNe and AGB stars are retained within the ISM and recycled into subsequent generations of stars. More importantly, the extended star formation histories permit a substantial contribution from Type Ia SNe. In our model, the minimum delay time for SNe Ia is 40 Myr, and roughly one quarter of all SNe Ia occur within the first 500 Myr. As a result, stars that form after reionization can undergo significantly stronger iron enrichment than those that formed earlier, as we saw in the m5.6 and m5.5 galaxies. 

The bottom panel of Fig.~\ref{fig:FeH_reionization} reveals a noticeable change in the character of chemical evolution following reionization, particularly in the two most massive galaxies. Prior to reionization, the metallicity of newly formed stars evolves in a highly stochastic manner, often exhibiting abrupt rises or drops associated with gas accretion, mergers, and bursty star formation. After reionization, however, the newly formed stellar populations follow a much tighter enrichment sequence. Once cosmological gas accretion is suppressed, chemical evolution becomes increasingly governed by the processing of a finite gas reservoir and the delayed release of metals from previously formed stars. The lower-mass galaxies also continue to enrich after reionization, although their enrichment histories exhibit substantially larger scatter owing to their smaller gas reservoirs and more stochastic star formation. Nevertheless, even these systems experience nearly an order-of-magnitude increase in metallicity before star formation is ultimately quenched. Reionization therefore does not halt chemical evolution immediately; instead, it initiates a prolonged phase of post-reionization enrichment whose duration depends on the ability of the halo to retain gas. The chemical properties of UFDs thus preserve a fossil record of their reionization histories: halos that retain gas for longer after reionization undergo more extended enrichment and produce larger populations of relatively metal-rich stars.

\section{Discussions and conclusions}
\label{sec:discuss}
\subsection{The need for an inhomogeneous cosmic reionization}
As discussed in Section~\ref{sec:sample}, even at $z=0$, the isolated UFDs account for 39\% of all $10^9\,\Msun$ halos. We have seen in the previous sections that the properties of these isolated UFDs are highly sensitive to their reionization and halo assembly history.

Previous simulations usually adopt a time-variable but spatially-uniform UV background \citep[e.g.,][]{2009ApJ...703.1416F,2020MNRAS.493.1614F:Faucher-Giguere}. In such UVB models, reionization typically starts very early, at $z\gtrsim10$. However, among the isolated halos we selected, three of them (m5.5, m4.4a, and m3.6) remain unaffected by any background radiation up to $z \sim 6.5$, as no photons reach these halos before that time. For the remainder of the halos, although photons from nearby low-mass systems arrive earlier, the ionized fraction remains close to zero until a strong I-front passes through at $z=5$--$6$, rapidly reionizing the halo (see Fig.~\ref{fig:reion_hist}). The only outlier is m4.5, which exhibits a plateau in the ionized fraction at $f_{\HII}\sim0.25$ for $\sim100\,\mathrm{Myr}$ prior to reionization. This behavior is driven locally by strong feedback associated with a starburst. The late and rapid reionization of these isolated UFDs reflects their environments: halos that remain isolated to $z=0$ tend to reside in relatively underdense regions, far from the strong ionizing sources that drive the early stages of reionization. Such environmentally dependent reionization histories cannot be captured by a spatially uniform UVB, highlighting the importance of modeling the inhomogeneous nature of cosmic reionization when studying the evolution of isolated UFDs. 


\subsection{Caveats}
\label{sec:cav}
We note that there are a few caveats and limitations to this work. Firstly, we neglect the contribution of the cosmic UV background in the Lyman--Werner (LW) band \citep[e.g.,][]{2023MNRAS.522..330I:Incatasciato}, which is not directly available from the \thesan simulation. LW photons, with energies in the range $11.2\text{--}13.6\,\mathrm{eV}$, are crucial for the photodissociation of molecular hydrogen. \cite{2026MNRAS.548ag439B:Brown} demonstrates that the intensity of the LW background shapes the halo occupation fraction and SHMR of UFDs, as it strongly influences the cooling processes and primordial chemistry of metal-poor gas.

Secondly, we do not adopt a dedicated model for the formation and evolution of Pop III stars. Instead, we form stars across all metallicities using the same set of criteria and the same IMF. We also do not consider  pair-instability SNe, and the metal yields of Pop III stars are derived by straightforwardly extrapolating the model used for Pop II stars. We have seen in Section~\ref{sec:MZR} that this could lead to the relatively low metallicity of the low-mass dwarf galaxies. However, it is unlikely to modify our results about the diversity of UFDs because it mainly impacts the early Pop~III-to-Pop~II transition \citep{2025ApJ...993....2B}. The formation and evolution of Pop III stars are still highly uncertain (see \citealt{2023ARA&A..61...65K:Klessen} for a review). One of the goals of the RIGEL project is to develop a subgrid Pop III star formation model that combines the analytical framework of \cite{2024MNRAS.534..290L:Liu,Gurian2025,Gurian2026} for Pop~III star formation with the Pop~III feedback prescription of \cite{2026arXiv260526206S:Saha}, integrating both within RIGEL. With this new model, we aim in future work to investigate a variety of Pop III formation and feedback scenarios and to predict how they influence the present-day properties of dwarf galaxies.

Moreover, dwarf galaxies typically do not host many massive molecular clouds or large clusters of massive stars, owing to their limited gas content and low SFR. Consequently, low-SFR dwarf galaxies may be deficient in massive stars relative to MW-like systems \citep[e.g.,][]{2009ApJ...706..599L:Lee,2026arXiv260610558S:Salvador}. In this work, however, we adopt a universal \cite{2003PASP..115..763C} IMF and sample stars via stochastic sampling. This assumption may substantially affect the stellar feedback and chemo-dynamical evolution of dwarf galaxies \citep[e.g.,][]{2014MNRAS.437.3980P:Ploeckinger}. Recently, \cite{2026arXiv260609999D:Deng} introduced a semi-deterministic IMF sampling scheme to form individual stars that can self-consistently account for IMF variations as a function of the galaxy SFR. We plan to implement this approach in future cosmological simulations of dwarf galaxies and compare the resulting models with those presented here.

\subsection{Summary}
In this paper, we have presented a suite of cosmological zoom-in simulations of eight isolated ultra-faint dwarf galaxies (UFDs) with present-day halo masses of $\sim10^9\,\Msun$. The simulations were performed with the RIGEL galaxy formation framework coupled to on-the-fly radiative transfer and realistic large-scale radiation boundary conditions extracted from the \thesan RHD simulation. This approach allows us to follow the interaction between UFDs and a spatially inhomogeneous, time-dependent radiation field during cosmic reionization while simultaneously resolving the multi-phase ISM at a mass resolution of $17.8\,\Msun$. Our main findings are summarized as follows:

\begin{itemize}
\item The simulated galaxies reproduce many observed properties of Local Group UFDs. All simulated galaxies are ancient stellar systems, with star formation confined to the first $\sim1.5$ Gyr of cosmic history and luminosity-weighted ages exceeding 12.6 Gyr, consistent with the observed old stellar populations of Local Group UFDs. Despite residing in nearly identical $z=0$ halos, the simulated galaxies exhibit a $\sim2$ dex spread in stellar mass, substantial variations in metallicity, and diverse star formation histories. 

\item The simulated galaxies broadly reproduce both the luminosity--size and luminosity--metallicity relations of Local Group dwarf galaxies. While the most luminous systems match the observed metallicities of UFDs, the faintest galaxies remain systematically metal-poor, likely reflecting the absence of dedicated Population III enrichment physics in our model. Nevertheless, our UFDs are significantly more metal-rich than those produced in many previous simulations, indicating that radiative feedback and explicit IMF sampling enable prolonged chemical enrichment and more efficient metal retention in low-mass galaxies.

\item All simulated UFDs are dispersion-supported and highly elongated ($e_\text{3D}>0.4$) systems. We argue that these large ellipticities arise naturally from bursty high-redshift assembly and are preserved by the early termination of star formation, suggesting that strongly elongated stellar distributions may be a generic property of isolated reionization relics.

\item Reionization quenches star formation through a two-stage process. The arrival of the I-front rapidly photoionizes and heats the diffuse neutral gas surrounding the galaxy, removing the reservoir that continuously replenishes the ISM. Star formation nevertheless continues for several hundred Myr using the surviving self-shielded gas. Quenching occurs only after this remaining gas is consumed or dispersed and can no longer cool and recollapse in the presence of the UV background.

\item Photoevaporation is the dominant channel through which baryons are removed from UFDs after reionization. Within 500\,Myr of reionization, less than 40\% of the initial gas mass remains within the Lagrangian volume of the halo, and most of this surviving gas resides in a diffuse ionized phase. The mass removed by photoevaporative flows substantially exceeds that carried by fast stellar-feedback-driven outflows.

\item The halo mass at the time of reionization, $M_{\rm halo,zion}$, is a key parameter governing the diversity of UFDs. Galaxies residing in more massive halos at reionization retain gas for longer periods, continue forming stars for a longer time after reionization, and ultimately produce larger stellar masses. We find strong correlations between $M_{200}^\text{reion}$ and the final stellar mass, quenching timescale, stellar age spread, and chemical enrichment history.

\item Reionization does not immediately terminate chemical evolution. In all simulated UFDs, stars continue to form and enrich the ISM after the arrival of the I-front. Even the lowest-mass systems experience nearly one dex of metallicity enrichment before star formation is fully quenched, while the most massive systems develop extended metal-rich stellar populations through prolonged post-reionization star formation.
\end{itemize}

Our results suggest that part of the observed diversity of Local Group UFDs may represent a fossil record of inhomogeneous reionization. Variations in the timing of reionization and in the halo mass at reionization translate into differences in stellar mass, star-formation duration, stellar age distribution, and chemical enrichment. In particular, the extent of post-reionization star formation and enrichment may provide an observable signature of the halo mass and evolutionary state at the time of reionization, which in turn depend on the interplay between halo assembly and the local timing of reionization.
Overall, we find that reionization is not merely a mechanism that suppresses star formation in low-mass halos, but a fundamental process that shapes the present-day diversity of UFDs. By coupling realistic patchy reionization to high-resolution galaxy formation simulations, we show that the properties of UFDs preserve valuable information about both their early assembly histories and the inhomogeneous nature of cosmic reionization.

\begin{acknowledgements}
We thank Volker Springel for giving us access to \arepo.  We use \textsc{python} packages {\sc NumPy} \citep{harris2020array}, {\sc SciPy} \citep{2020SciPy-NMeth}, {\sc astropy} \citep{2013A&A...558A..33A,2018AJ....156..123A}, {\sc matplotlib} \citep{Hunter:2007}, and {\sc Paicos} \citep{2024JOSS....9.6296B:Berlok} to analyze and visualize the simulation data. HL is supported by the National Key R\&D Program of China No. 2023YFB3002502, the National Natural Science Foundation of China under No. 12373006 and 12533004, and the China Manned Space Program with grant No. CMS-CSST-2025-A10. YD is supported by the National Natural Science Foundation of China under No. 125B2057. 
\end{acknowledgements}



\bibliographystyle{aa}
\bibliography{DwarfGalaxy} 

\begin{appendix}
\section{Definition of reionization redshift}
\label{sec:def_zion}
\begin{figure*}
	\includegraphics[width=2\columnwidth]{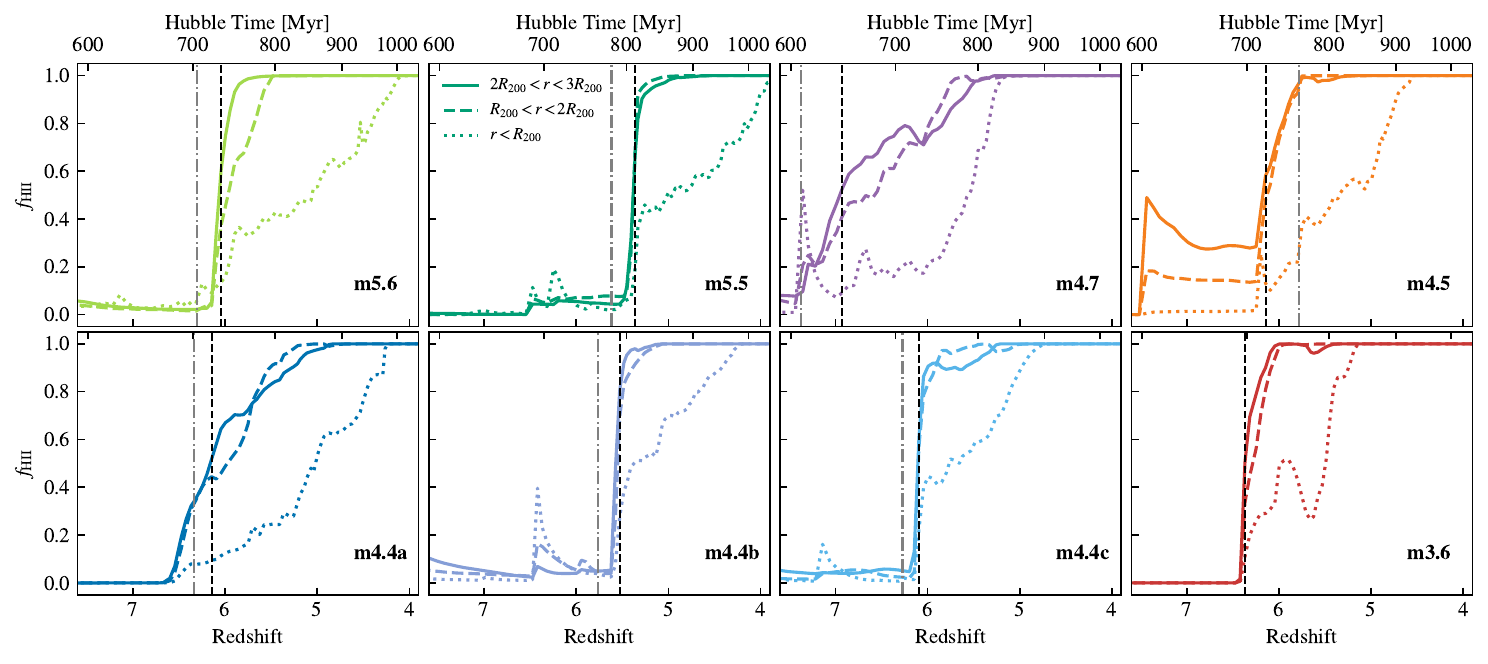}
    \caption{Evolution of the ionized gas fraction, $f_{\rm HII}$, within three radial shells surrounding each simulated UFD. Dotted, dashed, and solid curves show the ionized fraction inside $R_{200}$, between $R_{200}$ and $2R_{200}$, and between $2R_{200}$ and $3R_{200}$, respectively. The vertical dash-dotted line marks the reionization time in the \thesan simulation, defined as the time when the ionized fraction within a $(93.3\,\text{kpc})^3$ region enclosing the halo first exceeds 50\%. The vertical dashed line marks the time when the ionized fraction within the $2R_{200}<r<3R_{200}$ shell first exceeds 50\%.
    }
    \label{fig:reion_hist}
\end{figure*}
The timing of reionization varies substantially among the eight simulated UFDs, reflecting the patchy nature of cosmic reionization. Figure~\ref{fig:reion_hist} shows the evolution of the mass-weighted ionized gas fraction within three radial shells around each halo. The ionization histories differ significantly from halo to halo, with local reionization occurring over a redshift range of approximately $\Delta z\sim1.5$. 

To examine the impact of reionization on galaxy evolution, it is necessary to define a robust reionization epoch for each halo. In all halos, the gas outside the virial radius becomes ionized first, while the gas inside the halo responds more gradually owing to self-shielding and higher gas densities. The ionization histories in the $R_{200}<r<2R_{200}$ and $2R_{200}<r<3R_{200}$ shells are generally very similar, exhibiting a rapid transition from predominantly neutral to predominantly ionized gas over a timescale of several tens of Myr. This behavior indicates that both shells primarily trace the ionization state of the surrounding IGM and respond coherently to the passage of the large-scale I-front. 
In contrast, the gas within the virial radius often exhibits a more complex evolution. In several halos, the ionized fraction inside $R_{200}$ increases more gradually and displays significant fluctuations. This behavior arises from the presence of dense gas clumps and self-shielded structures that can remain neutral long after the surrounding IGM has become ionized. Consequently, the ionization history within the halo reflects not only the arrival of the I-front but also the subsequent photoevaporation and dispersal of dense gas.

Because our goal is to characterize the timing of the large-scale reionization process rather than the later evolution of self-shielded gas within halos, the ionization fraction measured inside $R_{200}$ does not provide a robust definition of the reionization epoch. We therefore define the reionization redshift of an individual halo, $z_\text{reion}$, as the time when the ionized fraction within the $2R_{200}<r<3R_{200}$ shell first exceeds 50\%. This definition traces the ionization state of the local IGM while remaining largely insensitive to dense gas structures associated with the halo itself. We also analyzed the evolution of the volume-weighted ionized gas fraction, which is used in \citep{Zhao2026} to define the reionization time, and found results broadly consistent with those based on the mass-weighted fraction. Only the m4.7 and m4.5 galaxies show earlier reionization times, by approximately 50 and 100\,Myr, respectively, while the differences for the other galaxies remain within 10\,Myr. However, these earlier reionization times do not accurately trace reionization by the background radiation, as supernova-driven bubbles can fill a large volume with highly ionized gas and thereby cause the volume-weighted ionized fraction to rise. We therefore adopt the mass-weighted ionized gas fraction to define the reionization time.

\section{Effects of the reduced speed of light approximation}
\label{sec:rsla}

\begin{figure}
	\includegraphics[width=\columnwidth]{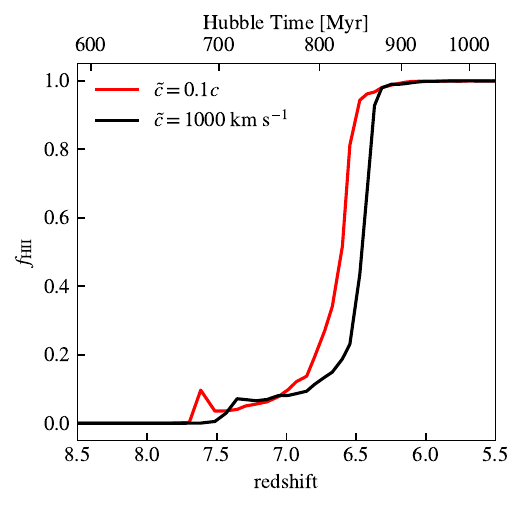}
    \caption{Evolution of the ionized gas fraction, $f_{\rm HII}$, within the $2R_{200}<r<3R_{200}$ shell for two different choices of the speed of light. The red curve shows the simulation adopting a reduced speed of light $\tilde{c}=0.1c$, while the black curve corresponds to our simulations using a speed of $1000\,{\rm km\,s^{-1}}$, comparable to the expected speed of I-fronts during reionization. 
    The epoch at which the shell becomes half ionized differs by $\Delta z_\text{reion}=0.17$ between the two runs, corresponding to a time difference of $\Delta t_\text{reion}=29$\,Myr.}
    \label{fig:rsla}
\end{figure}

A reasonable concern is that the local reionization time may be affected by
the reduced-speed-of-light approximation adopted in our simulations.
To assess this effect, we compare our simulation in which the speed of light is reduced to a fixed speed of $1000\,{\rm km\,s^{-1}}$, with a much faster speed of $\tilde{c}=0.1c$. Because the $0.1c$ run is very expensive with our high-resolution ICs, we instead perform this test using the m8.2 halo IC from \cite{2025OJAp....8E.153K}, which has eight times lower mass resolution.

The $\tilde{c}=0.1c$ run exhibits an earlier rise in $f_{\rm HII}$, whereas
the $1000\,{\rm km\,s^{-1}}$ run shows a delay.
The epoch at which the shell becomes half ionized differs by $\Delta z_\text{reion}=0.17$ between the two runs, corresponding to a time difference of $\Delta t_\text{reion}=29$\,Myr. This delay is substantially smaller than the $\sim300$\,Myr
difference in reionization times of the early and late reionized halos.
Therefore, while the reduced-speed-of-light approximation may alter the
detailed structure of I-fronts, it has little impact on the
determination of $z_\text{reion}$ as defined in this work. We conclude that
our measurements of local reionization times are robust against the choice
of radiation propagation speed.
\end{appendix}
\end{document}